\documentclass[aps,prb,twocolumn,showpacs,floatfix]{revtex4-1}
\usepackage{graphicx,amsfonts,amssymb,amsmath,hyperref}
\usepackage{textcomp}
\usepackage{esint}

\newif\ifhyper
\hypertrue
\ifhyper       
\hypersetup{         
  citecolor = {green},
  colorlinks = {true}, 
  urlcolor = {blue} 
}

\begin{document}

\graphicspath{{./figures_submit/}}

\def\rhoeq{\hat\rho_{\rm eq}}

\newcommand{\marge}[1]{\marginpar{\scriptsize #1}}
\newcommand{\remarque}[1]{\marginpar{\scriptsize Remarque}{\it [#1]}}
\newcommand{\new}[1]{{\bf #1}}
\newcommand{\red}[1]{\textcolor{red}{#1}}
\newcommand{\blue}[1]{\textcolor{blue}{#1}}
\newlength{\textlarg}
\newcommand{\barre}[1]{%
   \settowidth{\textlarg}{#1}
   #1\hspace{-\textlarg}\rule[0.5ex]{\textlarg}{0.5pt}}
\newcommand{\barred}[1]{%
   \settowidth{\textlarg}{#1}
   \red{#1}\hspace{-\textlarg}\rule[0.5ex]{\textlarg}{0.5pt}}
\newcommand{\barblue}[1]{%
   \settowidth{\textlarg}{#1}
   \blue{#1}\hspace{-\textlarg}\rule[0.5ex]{\textlarg}{0.5pt}}

\def\beq{\begin{equation}}
\def\eeq{\end{equation}}
\def\bleq{\begin{eqnarray}}
\def\eleq{\end{eqnarray}} 
\def\bfig{\begin{figure}}
\def\efig{\end{figure}}
\def\bline{\begin{multline}}
\def\eline{\end{multline}}
\def\bremark{\begin{quotation} \noindent \small }
\def\eremark{\end{quotation}}
\def\llbrace{\left\lbrace}
\def\rrbrace{\right\rbrace}
\def\lbraket{\left[}
\def\rbraket{\right]}
\def\llangle{\left\langle}
\def\rrangle{\right\rangle} 

\newcommand{\Tr}{{\rm Tr}} 
\newcommand{\tr}{{\rm tr}} 
\newcommand{\sgn}{{\rm sgn}} 
\newcommand{\mean}[1]{\langle #1 \rangle}
\newcommand{\commu}[2]{[#1,#2]} 
\newcommand{\bra}[1]{\langle#1|}
\newcommand{\ket}[1]{|#1\rangle}
\newcommand{\braket}[2]{\langle #1|#2\rangle}
\newcommand{\dbraket}[3]{\langle #1|#2|#3\rangle}
\newcommand{\tens}[1]{\overleftrightarrow{#1}}  
\newcommand{\vac}{|{\rm vac}\rangle} 
\def\bravac{\langle{\rm vac}|}
\newcommand{\const}{{\rm const}} 
\newcommand{\atanh}{\,{\rm atanh}}

\newcommand{\ie}{i.e. }
\newcommand{\iet}{i.e.}
\newcommand{\eg}{e.g. }
\newcommand{\cc}{{\rm c.c.}} 
\newcommand{\hc}{{\rm h.c.}} 
\def\etal{{\it et al. }}

\newcommand{\jhatbf}{\hat {\textbf \j}} 
\newcommand{\Jhatbf}{\hat {\textbf \J}} 
\newcommand{\jhat}{\hat {\jmath}} 
\newcommand{\Jhat}{\hat {J}} 
\newcommand{\jbf}{\textbf j}
\newcommand{\Jbf}{\textbf J}

\def\chibf{\boldsymbol{\chi}}
\def\down{\downarrow}
\def\eps{\epsilon}
\def\gam{\gamma} 
\def\phibf{\boldsymbol{\phi}}
\def\varphibf{\boldsymbol{\varphi}}
\def\varphibfs{\boldsymbol{\varphi}_<}
\def\varphibfl{\boldsymbol{\varphi}_>}
\def\varphis{\varphi_{<}}
\def\varphil{\varphi_{>}}
\def\psibf{\boldsymbol{\psi}}
\def\Ome{\Omega}
\def\omeD{{\omega_D}} 
\def\bfOme{\boldsymbol{\Omega}} 
\def\Omebf{\boldsymbol{\Omega}} 
\def\lamb{\lambda}
\def\Lamb{\Lambda}
\def\sig{\sigma}
\def\Sig{\Sigma}
\def\sigp{{\sigma'}} 
\def\bfsig{\boldsymbol{\sigma}} 
\def\sigbf{\boldsymbol{\sigma}} 
\def\The{\Theta} 
\def\up{\uparrow}

\def\epsk{\epsilon_{\bf k}} 
\def\xik{\xi_{\bf k}} 
\def\txik{\tilde\xi_{\bf k}} 
\def\xip{\xi_{\bf p}} 
\def\xikq{\xi_{{\bf k}+{\bf q}}} 
\def\Ek{E_{\bf k}} 
\def\Ep{E_{\bf p}}
\def\Heff{\hat H_{\rm eff}}
\def\Hem{\hat H_{\rm em}}
\def\Hint{\hat H_{\rm int}}
\def\Hloc{\hat H_{\rm loc}}
\def\HMF{\hat H_{\rm MF}}
\def\Sem{S_{\rm em}}
\def\SMF{S_{\rm MF}} 
\def\SHF{S_{\rm HF}} 
\def\SRPA{S_{\rm RPA}} 
\def\Sint{S_{\rm int}} 
\def\Sloc{S_{\rm loc}}
\def\TN{T_{\rm N}} 
\def\TNHF{T^{\rm HF}_{\rm N}} 
\def\Zloc{Z_{\rm loc}} 
\def\ZMF{Z_{\rm MF}} 
\def\ZHF{Z_{\rm HF}} 
\def\ZRPA{Z_{\rm RPA}} 
\def\RPA{{\rm RPA}}
\def\loc{{\rm loc}} 
\def\pp{{\rm pp}}
\def\ph{{\rm ph}} 
\def\ch{{\rm ch}}
\def\sp{{\rm sp}} 
\def\qtf{q_{\rm TF}}
\def\epstf{\eps^{}_{\rm TF}} 
\def\epsrpa{\eps^{}_{\rm RPA}} 
\def\chinnzpp{\chi_{nn}^{0}{}\!\!\!''}

\def\half{\frac{1}{2}}
\def\dhalf{\dfrac{1}{2}}
\def\third{\frac{1}{3}} 
\def\quarter{\frac{1}{4}}

\def\qr{{\bf q}\cdot{\bf r}}
\def\wt{\omega t} 

\def\a{{\bf a}}
\def\b{{\bf b}}
\def\e{{\bf e}}
\def\f{{\bf f}}
\def\g{{\bf g}}
\def\h{{\bf h}}
\def\k{{\bf k}}
\def\l{{\bf l}}
\def\m{{\bf m}}
\def\n{{\bf n}} 
\def\p{{\bf p}} 
\def\q{{\bf q}}
\def\r{{\bf r}}
\def\t{{\bf t}}
\def\u{{\bf u}}
\def\v{{\bf v}}
\def\x{{\bf x}}
\def\y{{\bf y}} 
\def\z{{\bf z}} 
\def\A{{\bf A}}
\def\B{{\bf B}}
\def\D{{\bf D}} 
\def\E{{\bf E}} 
\def\F{{\bf F}} 
\def\H{{\bf H}}  
\def\J{{\bf J}}
\def\K{{\bf K}} 

\def\G{{\bf G}}
\def\L{{\bf L}}
\def\M{{\bf M}}  
\def\O{{\bf O}} 
\def\P{{\bf P}} 
\def\Q{{\bf Q}} 
\def\R{{\bf R}}
\def\S{{\bf S}}
\def\epsbf{\boldsymbol{\epsilon}}
\def\mubf{\boldsymbol{\mu}}
\def\nablabf{\boldsymbol{\nabla}}
\def\rhobf{\boldsymbol{\rho}}
\def\sigmabf{\boldsymbol{\sigma}} 
\def\Pibf{\boldsymbol{\Pi}}
\def\pibf{\boldsymbol{\pi}}

\def\para{\parallel}
\def\kpara{{k_\parallel}}
\def\kperp{{k_\perp}} 
\def\kperpp{{k_\perp'}} 
\def\qperp{{q_\perp}} 
\def\tperp{{t_\perp}} 

\def\w{\omega}
\def\wn{\omega_n}
\def\wnu{\omega_\nu}
\def\wp{\omega_p} 
\def\dmu{{\partial_\mu}}
\def\dl{{\partial_l}}  
\def\dt{\partial_t} 
\def\tdt{\tilde\partial_t}
\def\dk{\partial_k}
\def\tdk{\tilde\partial_k}
\def\dx{\partial_x}
\def\dy{\partial_y} 
\def\dtau{{\partial_\tau}}  
\def\det{{\rm det}} 
\def\Pf{{\rm Pf}}

\def\dsum{\displaystyle \sum}
\def\dint{\displaystyle \int} 
\def\intt{\int_{-\infty}^\infty dt} 
\def\inttp{\int_{-\infty}^\infty dt'} 
\def\intk{\int_{\bf k}} 
\def\intkd{\int \frac{d^dk}{(2\pi)^d}}
\def\intq{\int_{\bf q}} 
\def\intr{\int d^dr}  
\def\dintr{\displaystyle \int d^dr} 
\def\intrp{\int d^dr'}
\def\dinttau{\displaystyle \int_0^\beta d\tau}
\def\dinttaup{\displaystyle \int_0^\beta d\tau'}
\def\inttau{\int_0^\beta d\tau}
\def\inttaup{\int_0^\beta d\tau'}
\def\intx{\int d^{d+1}x} 
\def\inttaur{\int_0^\beta d\tau \int d^dr}
\def\intinf{\int_{-\infty}^\infty}
\def\dinttaur{\displaystyle \int_0^\beta d\tau \int d^dr}
\def\dintinf{\displaystyle \int_{-\infty}^\infty}
\def\intw{\int_{-\infty}^\infty \frac{d\w}{2\pi}}
\def\sumr{\sum_{\bf r}} 

\def\calA{{\cal A}}
\def\calB{{\cal B}} 
\def\calC{{\cal C}} 
\def\dt{\partial_t}
\def\calD{{\cal D}}
\def\calF{{\cal F}} 
\def\calG{{\cal G}}
\def\calH{{\cal H}}
\def\calI{{\cal I}}
\def\calJ{{\cal J}}
\def\calK{{\cal K}}
\def\calL{{\cal L}} 
\def\calN{{\cal N}}
\def\calO{{\cal O}}
\def\calP{{\cal P}}  
\def\calR{{\cal R}} 
\def\calS{{\cal S}}
\def\calT{{\cal T}}
\def\calU{{\cal U}}
\def\calX{{\cal X}} 
\def\calY{{\cal Y}} 
\def\calZ{{\cal Z}} 

\def\calFbf{{\bf F}}

\def\tT{{\tilde T}}
\def\talpha{{\tilde\alpha}}
\def\tdelta{{\tilde\delta}}
\def\teta{{\tilde\eta}} 
\def\tlamb{{\tilde\lambda}}
\def\tmu{{\tilde\mu}}
\def\tphibf{{\tilde\phibf}}
\def\trho{{\tilde\rho}}
\def\tvarphibf{{\tilde\varphibf}} 
\def\tw{{\tilde\omega}}
\def\twn{{\tilde\omega_n}}

\def\asinh{{\rm asinh}} 

\def\Tkt{T_{\rm BKT}}
\def\bdelta{\bar\delta} 
\def\GamA{\Gamma_{A,k}}
\def\GamB{\Gamma_{B,k}}
\def\GamBL{\Gamma_{B,\Lambda}}
\def\DA{\Delta_{A,k}}
\def\DB{\Delta_{B,k}}
\def\YA{Y_{A,k}}
\def\YB{Y_{B,k}}

\title{Higgs amplitude mode in the vicinity of a $(2+1)$-dimensional quantum critical point: a nonperturbative renormalization-group approach}  

\author{F. Rose}
\author{F. L\'eonard}
\author{N. Dupuis}
\affiliation{Laboratoire de Physique Th\'eorique de la Mati\`ere Condens\'ee, 
CNRS UMR 7600, Universit\'e Pierre et Marie Curie, 4 Place Jussieu, 
75252 Paris Cedex 05, France}

\date{May 29, 2015} 

\begin{abstract}
We study the ``Higgs'' amplitude mode in the relativistic quantum O($N$) model in two space dimensions. Using the nonperturbative renormalization group and the Blaizot--M\'endez-Galain--Wschebor approximation (which we generalize to compute 4-point correlation functions), we compute the O($N$) invariant scalar susceptibility  at zero temperature in the vicinity of the quantum critical point. In the ordered phase, we find a well-defined Higgs resonance for $N=2$ and $N=3$ and determine its universal properties.  No resonance is found for $N\geq 4$. In the disordered phase, the spectral function exhibits a threshold behavior with no Higgs-like peak. We also show that for $N=2$ the Higgs mode manifests itself as a very broad peak in the longitudinal susceptibility in spite of the infrared divergence of the latter. We compare our findings with results from quantum Monte Carlo simulations and $\eps=4-(d+1)$ expansion near $d=3$. 
\end{abstract}
\maketitle 


\section{Introduction}

Relativistic quantum field theories with O($N$) symmetry arise in the low-energy description of many condensed-matter systems: quantum antiferromagnets, superconductors, Bose-Einstein condensates in optical lattices, etc. In the ordered phase, where the O($N$) symmetry is spontaneously broken, mean-field theory predicts $N-1$ gapless Goldstone modes corresponding to fluctuations of the direction of the $N$-component quantum field, and a gapped amplitude ``Higgs'' mode (see, for instance, Ref.~\onlinecite{Sachdev_book}).  

The existence of the Higgs mode near the quantum critical point (QCP) separating the ordered and disordered phases, when fluctuations are taken into account beyond mean-field theory, has been a subject of debate. Does the Higgs mode exist as a resonance-like feature or is it overdamped due to its coupling to the Goldstone modes? Space dimensionality $d$ plays a crucial role. In three dimensions, the QCP corresponds to a Gaussian fixed point of the renormalization group and interactions are suppressed at low energies; the Higgs resonance becomes sharper and sharper as the QCP is approached. This has been beautifully confirmed in the quantum antiferromagnet TlCuCl$_3$.\cite{Ruegg08,[{See also Ref. }] [{ for an experiment with cold atoms.}] Bissbort11} By contrast, for $2<d+1<4$, the QCP corresponds to the Wilson-Fisher fixed point and interactions are strong at low energies; the existence of a well-defined Higgs resonance is not guaranteed. Furthermore the visibility of the Higgs mode strongly depends on the 
symmetry of the probe.\cite{Podolsky11} For $2<d+1<4$, emission of Goldstone bosons leads to an infrared divergence in the longitudinal susceptibility,\cite{Patasinskij73,Sachdev99,Zwerger04,Dupuis11} which is the standard correlation function to probe the amplitude mode, thus making the observation of the Higgs resonance very difficult. The O($N$)-invariant scalar susceptibility (i.e. the correlation function of the square of the order parameter field) has a spectral weight which vanishes at low energies and is a much better candidate.\cite{Podolsky11} The Higgs mode in a two-dimensional system has been observed in a Bose gas in an optical lattice in the vicinity of the superfluid--Mott-insulator transition\cite{Endres12} and in a disordered superconductor close to the superconductor-insulator transition.\cite{Sherman15} 

The Higgs mode near a two-dimensional relativistic QCP has been studied with various techniques: large-$N$ expansion,\cite{Podolsky12} quantum Monte Carlo simulations,\cite{Pollet12,Chen13,Gazit13,Gazit13a} nonperturbative renormalization group (NPRG),\cite{Rancon14} and $\eps=4-(d+1)$ expansion about $d=3$.\cite{Katan15} For $N=2$, these studies have conclusively shown the existence of a Higgs resonance in the ordered phase which persists arbitrary close to the QCP. However, besides quantitative issues, such as the precise value of the mass of the Higgs mode, some basic qualitative questions remain: is there a Higgs-like resonance also in the disordered phase? Does the Higgs resonance exist for larger values of $N$, e.g. $N=3$ and $N=4$? 

In this paper we address these issues using the NPRG approach.\cite{Berges02,Delamotte12,Kopietz_book} More specifically, we use the Blaizot--M\'endez-Galain--Wschebor (BMW) approach, an approximation scheme to the exact RG equations which allows one to obtain the momentum and frequency dependence of correlation functions.\cite{Blaizot06,Benitez09,Benitez12,Parola84,Parola95} The BMW approximation is also the starting point of the NPRG study reported in Ref.~\onlinecite{Rancon14}. In that work, however, additional (uncontrolled) approximations were made in order to simplify the numerical solution of the NPRG-BMW equations. 

The main results of our analysis, which is restricted to zero temperature, are the following. i) In the ordered phase, we find a well-defined Higgs resonance for $N=2$ in agreement with previous works.\cite{Pollet12,Chen13,Gazit13,Gazit13a,Rancon14,Katan15} ii) We show that a Higgs resonance is also present for $N=3$ but not for $N=4$. This differs from previous NPRG analysis\cite{Rancon14} but agrees with Monte Carlo (MC) results.\cite{Gazit13,Gazit13a} iii) In the disordered phase we find that the spectral function of the scalar susceptibility does not exhibit a Higgs-like resonance peak above the absorption threshold. This contradicts previous NPRG results\cite{Rancon14} and some QMC simulations.\cite{Chen13} On the other hand, it is corroborated by a separate MC analysis,\cite{Gazit13a} which found no conclusive evidence for such a resonance, and is supported by the results of the $\eps=4-(d+1)$ expansion. iv) Although the longitudinal susceptibility diverges as $1/\w$ at low energies due to its 
coupling to the Goldstone modes, for $N=2$ the Higgs mode manifests itself in the spectral function as a very broad peak. 

The outline of the paper is as follows. In Sec.~\ref{sec_nprg} we present the NPRG approach to the quantum O($N$) model at zero temperature. In that limit, the two-dimensional quantum model is equivalent to the three-dimensional classical model. We show that the BMW approximation can be extended to compute 4-point correlation functions such as the scalar susceptibility. We also recall the expected scaling behavior of the longitudinal, transverse and scalar susceptibilities near the QCP. In Sec.~\ref{sec_largeN} we show that the BMW approach becomes exact in the limit $N\to\infty$. The  numerical solution of the RG equations for $N\gg 1$ and $N=2,3,4,\cdots$ is discussed in Sec.~\ref{sec_spectral}. Both the longitudinal and scalar susceptibilities are considered. We compute the spectral functions and the corresponding universal scaling functions, as well as the mass of the Higgs mode and the stiffness. In Sec.~\ref{sec_conlu} we provide a summary and conclusion.

\section{NPRG approach}
\label{sec_nprg} 

The two-dimensional quantum O($N$) model is defined by the (Euclidean) action 
\begin{align}
S[\varphibf] ={}& \int_\x \biggl\lbrace \half (\nablabf\varphibf)^2 + \frac{1}{2c^2} (\dtau \varphibf)^2 \nonumber \\ & 
+ \frac{r_0}{2} \varphibf^2 + \frac{u_0}{4!N} (\varphibf^2)^2  \biggr\rbrace, 
\label{action1} 
\end{align}
where we use the shorthand notation 
\begin{equation}
\x=(\r,\tau), \quad \int_\x = \inttau \int d^2r .
\end{equation}
$\varphibf(\x)$ is an $N$-component real field, $\tau\in [0,\beta]$ an imaginary time and $\r$ a two-dimensional coordinate ($\beta=1/T$ and we set $\hbar=k_B=1$). $r_0$ and $u_0$ are temperature-independent coupling constants and $c$ is the (bare) velocity of the $\varphibf$ field. The factor $1/N$ in Eq.~(\ref{action1}) is introduced to obtain a meaningful limit $N\to\infty$ (with $u_0$ fixed).
The model is regularized by an ultraviolet cutoff $\Lambda$. In order to maintain the Lorentz invariance of the action~(\ref{action1}) at zero temperature, it is natural to implement a cutoff on both momenta and frequencies.

The phase diagram of the quantum O($N$) model with $N\geq 2$ is well known. At zero temperature, there is a quantum phase transition between a disordered phase ($r_0>r_{0c}$) and an ordered phase ($r_0<r_{0c}$) where the O($N$) symmetry of the action~(\ref{action1}) is spontaneously broken ($u_0$ and $c$ are considered as fixed parameters). The QCP at $r_0=r_{0c}$ is in the universality class of the three-dimensional classical O($N$) model with a dynamical critical exponent $z=1$ (this value follows from Lorentz invariance); the phase transition is governed by the three-dimensional Wilson-Fisher fixed point. At finite temperatures, the system is always disordered for $N\geq 2$, in agreement with the Mermin-Wagner theorem. For $N=2$ and $r_0 < r_{0c}$, there is a finite-temperature Berezinskii-Kosterlitz-Thouless (BKT) phase transition\cite{Berezinskii70,*Berezinskii71,Kosterlitz73,Kosterlitz74} and the system exhibits algebraic order at low temperatures. The BKT transition temperature line $\Tkt$ terminates 
at the QCP $r_0=r_{0c}$. 

In the following, we consider only the zero temperature limit where the two-dimensional quantum O($N$) model is equivalent to the three-dimensional classical O($N$) model. For convenience, we set the velocity $c$ equal to one so that the action~(\ref{action1}) takes the usual form of the classical O($N$) model with $\x$ a three-dimensional space variable. Having in mind the two-dimensional quantum O($N$) model, we shall refer to the critical point of the three-dimensional classical O($N$) as the QCP. In Fourier space, a correlation function $\chi(p_x,p_y,p_z)$ computed in the classical model should be identified with the correlation function $\chi(p_x,p_y,i\w)$ in the quantum model (with $\w$ a bosonic Matsubara frequency) and yields the retarded dynamical correlation function $\chi^R(p_x,p_y,\w)\equiv \chi(p_x,p_y,i\w\to\w+i0^+)$ after analytical continuation $i\w\to \w+i0^+$. In the following, we shall often use the notation $\chi^R(\w)$ for $\chi^R(p_x=p_y=0,\w)$ and $\chi''(\w)={\rm Im}[\chi^R(\w)]$. 

\subsection{Scale-dependent effective action $\Gamma_k$}
\label{subsec_gamk} 

To implement the NPRG approach, we add to the action an infrared regulator term,
\begin{equation}
\Delta S_k[\varphibf] = \half \sum_{\q,i} \varphi_i(-\q) R_k(\q) \varphi_i(\q) ,
\label{DSk}
\end{equation}
such that fluctuations are smoothly taken into account as the momentum scale $k$ varies from the ultraviolet cutoff $\Lambda$ down to 0.\cite{Berges02,Delamotte12,Kopietz_book} The cutoff function in~(\ref{DSk}) is defined by 
\begin{equation}
R_k(\q) = Z_k \q^2 r\left(\frac{\q^2}{k^2}\right) , \quad r(y) = \frac{\alpha}{e^y-1} , 
\label{Rkdef}
\end{equation} 
where $\alpha$ is a constant of order one (we choose $\alpha=2.25$) and $Z_k$ a field renormalization factor defined in Sec.~\ref{subsec_explicit_rg}.\cite{not5} Thus the $\Delta S_k$ term suppresses fluctuations with momenta $|\q|\lesssim k$ but leaves unaffected those with $|\q|\gtrsim k$. 
We consider the $k$-dependent partition function 
\begin{equation}
Z_k[\J,h] = \int\calD[\varphibf]\, e^{-S[\varphibf]-\Delta S_k[\varphibf] + \int_\x (\J\cdot \varphibf + h\varphibf^2) } , 
\label{Zk} 
\end{equation}
in the presence of external sources $\J$ and $h$. The mean value of the field is defined by 
\begin{equation}
\phibf_k[\x;\J,h] = \frac{\delta\ln Z_k[\J,h]}{\delta\J(\x)} = \mean{\varphibf(\x)} . 
\label{phidef}
\end{equation}
The bilinear source $h$ will allow us to compute the scalar susceptibility (but will eventually be set to zero).

The central quantity in the NPRG approach is the scale-dependent effective action 
\begin{equation}
\Gamma_k[\phibf,h] = - \ln Z_k[\J,h] + \int_\x  \J\cdot\phibf - \Delta S_k[\phibf] ,
\label{gammak}
\end{equation}
defined as a modified Legendre transform of $-\ln Z_k[\J,h]$, wrt the linear source $\J$, which includes the subtraction of $\Delta S_k[\phibf]$. In Eq.~(\ref{gammak}), $\J(\x)\equiv\J_k[\x;\phibf,h]$ is obtained by inverting Eq.~(\ref{phidef}). Assuming that for $k=\Lambda$ the fluctuations are completely suppressed by the $\Delta S_k$ term, we have 
\begin{equation}
\Gamma_\Lambda[\phibf,h]=S[\phibf] - \int_\x h \phibf^2. 
\end{equation}
On the other hand, the effective action of the original model~(\ref{action1}) is given by $\Gamma_{k=0}$ since $R_{k=0}$ vanishes. 
The NPRG approach aims at determining $\Gamma_{k=0}$ from $\Gamma_\Lambda$ using Wetterich's equation\cite{Wetterich93}
\begin{equation}
\dk \Gamma_k[\phibf,h] = \half \Tr\llbrace \dot R_k\left(\Gamma^{(2,0)}_k[\phibf,h] + R_k\right)^{-1} \rrbrace ,
\label{rgeq}
\end{equation}
where $\dot R_k=\dk R_k$ and $\Gamma^{(2,0)}_k[\phibf,h]$ denotes the second-order functional derivative of $\Gamma_k[\phibf,h]$ with respect to $\phibf$. In Fourier space, the trace involves a sum over momenta as well as the O($N$) index of the $\phibf$ field.

Most of the physically relevant information encoded in $\Gamma_k$ can be obtained either from the effective potential or the one-particle-irreducible (1PI) vertices. The effective potential is nothing but $\Gamma_k$ evaluated in a constant, i.e. uniform, field configuration (up to a volume factor $V$),
\begin{equation}
U_k(\rho) = \frac{1}{V} \Gamma_k[\phi]\Bigl|_{\phibf=\const} . 
\end{equation}
Because of the O($N$) symmetry of $\Gamma_k$, $U_k$ is a function of the O($N$) invariant $\rho=\phibf^2/2$. We denote by $\rho_{0,k}$ the value of $\rho$ at the minimum of the effective potential. Spontaneous symmetry breaking of the O($N$) symmetry is characterized by a nonvanishing expectation value of the field $\varphibf$, i.e. $\lim_{k\to 0}\rho_{0,k}=\rho_{0}>0$. In this case, the effective potential $U_{k=0}(\rho)$ is constant (and minimum) for $0\leq \rho\leq\rho_0$ in agreement with the convexity of the Legendre transform $\Gamma_{k=0}[\phibf]$. At the QCP ($r_0=r_{0c}$), $\rho_0=0$ but $\rho_{0,k}>0$ for all nonzero values of $k$. 

The 1PI vertices are defined by 
\begin{multline}
\Gamma_{k,\lbrace i_j\rbrace}^{(n,m)}[\lbrace \x_j\rbrace,\lbrace \y_j\rbrace;\phibf,h] = \\ \frac{\delta^{n+m} \Gamma_k[\phibf,h]}{\delta \phi_{i_1}(\x_1)\cdots \delta \phi_{i_n}(\x_n) \delta h(\y_1) \cdots \delta h(\y_m)}  .
\label{1PIdef} 
\end{multline}
The correlation functions evaluated for $h=0$ and in a uniform field configuration are determined by the vertices 
\begin{equation}
\Gamma_{k,\lbrace i_j\rbrace}^{(n,m)}(\lbrace \x_j\rbrace,\lbrace \y_j\rbrace;\phibf) =
\Gamma_{k,\lbrace i_j\rbrace}^{(n,m)}[\lbrace \x_j\rbrace,\lbrace \y_j\rbrace;\phibf,h] \Bigl|_{\phibf=\const \atop h=0} .
\end{equation}
In particular, the propagator $G_{k,ij}(\p,\phibf)=\mean{\varphi_i(\p)\varphi_j(-\p)}_{h=0}$ in a uniform field 
is obtained from the matrix equation 
\begin{equation}
G_k(\p,\phibf) = \bigl( \Gamma_k^{(2,0)}(\p,\phibf)+R_k \bigr)^{-1} ,
\end{equation}
where $\Gamma_k^{(2,0)}(\p,\phibf)\equiv \Gamma_k^{(2,0)}(\p,-\p,\phibf)$.
The O($N$) symmetry allows us to write 
\begin{equation}
\Gamma^{(2,0)}_{k,ij}(\p,\phibf) = \delta_{i,j} \Gamma_{A,k}(\p,\rho) + \phi_i \phi_j \Gamma_{B,k}(\p,\rho) , 
\label{GamAB}
\end{equation}
which yields the longitudinal (L) and transverse (T) parts of the propagator, 
\begin{align}
G_{k,\rm L}(\p,\rho) &= [\Gamma_{A,k}(\p,\rho) + 2 \rho \Gamma_{B,k}(\p,\rho)+ R_k(\p)]^{-1} , \nonumber \\ 
G_{k,\rm T}(\p,\rho) &= [\Gamma_{A,k}(\p,\rho) + R_k(\p)]^{-1} .
\label{GLT} 
\end{align}
Due to rotation invariance in space, 2-point vertices and correlation functions in~(\ref{GamAB}) and (\ref{GLT}) are functions of $p=|\p|$.

Important information can be obtained from the longitudinal and transverse susceptibilities that we define by 
\begin{equation}
\chi_{\alpha}(\p)=G_{k=0,\alpha}(\p,\rho_{0,k=0}) \qquad (\alpha={\rm L,T}). 
\end{equation}
In the disordered phase ($\rho_0=0$), $\chi_{\rm L}(\p)=\chi_{\rm T}(\p)\equiv\chi_{\rm L,T}(\p)$. The single-particle excitation gap $\Delta$ is obtained from the vanishing of the spectral function $\chi''_{\rm L,T}(\p,\w)={\rm Im}[\chi^R_{\rm L,T}(\p,\w)]$ for $|\w|<\Delta$. The excitation gap manifests itself as a sharp peak in $\chi''_{\rm L,T}(\w)\equiv\chi''_{\rm L,T}(\p=0,\w)$ (see Sec.~\ref{subsubsec_disordered}). In the ordered phase, the stiffness $\rho_s$ is defined by\cite{Chaikin_book} 
\begin{equation}
\chi_{\rm T}(\p) =  \frac{2\rho_0}{\rho_s \p^2} \quad \mbox{for} \quad \p \to 0 . 
\label{rhosdef}
\end{equation}
For two systems located symmetrically wrt the QCP (i.e. corresponding to the same value of $|r_0-r_{0c}|$), one in the ordered phase (with stiffness $\rho_s$) and the other in the disordered phase (with excitation gap $\Delta$), the ratio $\rho_s/\Delta$ is a universal number which depends only on $N$. This allows us to use $\Delta$ as the characteristic energy scale in both the disordered and ordered phases (in the latter case, $\Delta$ is defined as the excitation gap at the point located symmetrically wrt the QCP).\cite{Podolsky12} Note that $\Delta$ and $\rho_s$ vanish as $|r_0-r_{0c}|^{\nu}$ as we approach the QCP. 

In the universal regime near the QCP (scaling limit),\cite{Sachdev_book}
\begin{equation}
\begin{split}
\chi_\alpha(\p) &= Z_{\alpha,\pm} \Delta^{\eta-2} \tilde\Phi_{\alpha,\pm} \left( \frac{p}{\Delta} \right) , \\
\chi''_\alpha(\w) & = {\rm Im}[\chi^R_\alpha(\w)] = Z_{\alpha,\pm} \Delta^{\eta-2} \Phi_{\alpha,\pm} \left( \frac{\w}{\Delta} \right) ,
\end{split}
\label{chiLT_scaling}
\end{equation}
where $\eta$ is the anomalous dimension of the $\varphibf$ field at the QCP. 
$\tilde\Phi_{\alpha,\pm}$ and $\Phi_{\alpha,\pm}$ are universal scaling functions and $Z_{\alpha,\pm}$ a nonuniversal constant with dimension of (length)$^\eta$. The index $+/-$ refers to the disordered and ordered phases, respectively. At the QCP ($\Delta=0$), $\chi_\alpha(\p)\sim p^{\eta-2}$ and $\chi''_\alpha(\w)\sim|\w|^{\eta-2}$. Since $\chi''_{\rm L}(\w)$ and $\chi''_{\rm T}(\w)$ are odd in $\w$ we shall only consider the case $\w\geq 0$ in the following.

\subsection{Scalar susceptibility}

\begin{figure}
\centerline{\includegraphics[width=7.3cm]{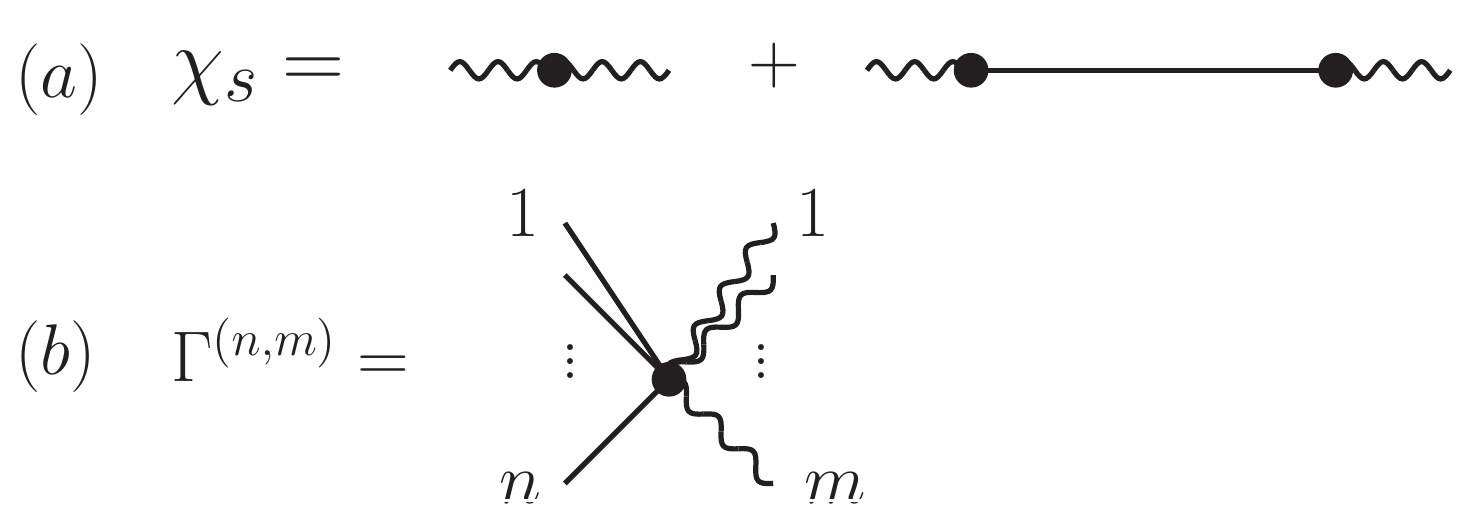}}
\caption{Diagrammatic representation of the scalar susceptibility~(\ref{chi5}) (a). A vertex $\Gamma^{(n,m)}$ is represented by a black dot with $n$ solid lines and $m$ wavy lines (b). Solid lines connecting vertices stand for the propagator $G=\Gamma^{(2,0)-1}$.} 
\label{fig_chis_diag} 
\end{figure}

We now consider the scalar susceptibility
\begin{align}
\chi_s(\y-\y') &= \mean{\varphibf(\y)^2 \varphibf(\y')^2} - \mean{\varphibf(\y)^2}\mean{\varphibf(\y')^2} \nonumber \\
&= \frac{\delta^2 \ln Z[\J,h]}{\delta h(\y) \delta h(\y')} \biggl|_{\J=h=0} ,
\label{chi1} 
\end{align}
where $Z[\J,h]\equiv Z_{k=0}[\J,h]$. Using~(\ref{gammak}), we can express $\chi_s$ as a functional derivative of the effective action $\Gamma\equiv\Gamma_{k=0}$, 
\begin{equation}
\chi_s(\y,\y') = - \frac{\bdelta^2 \Gamma[\bar\phibf[h],h]}{\bdelta h(\y) \bdelta h(\y')} \biggl|_{h=0} ,
\label{chi2} 
\end{equation}
where the order parameter $\bar\phibf[h]$ is defined by 
\begin{equation}
\frac{\delta\Gamma[\phibf,h]}{\delta\phibf(\x)} \biggl|_{\phibf=\bar\phibf[h]} = 0. 
\label{phibardef}
\end{equation}
In Eq.~(\ref{chi2}), $\bdelta/\bdelta h(\y)$ is a total derivative which acts both on $\bar\phibf[h]$ and the explicit $h$-dependence of the functional $\Gamma[\phibf,h]$. Using~(\ref{phibardef}), one finds 
\begin{multline}
\frac{\bdelta^2 \Gamma[\bar\phibf[h],h]}{\bdelta h(\y) \bdelta h(\y')} = \Gamma^{(0,2)}[\y,\y';\bar\phibf[h],h] \\
+ \int_\x \Gamma^{(1,1)}_i[\x,\y;\bar\phibf[h],h]  \frac{\delta \bar\phi_i[\x;h]}{\delta h(\y')}  
\label{chi3} 
\end{multline} 
(we use Einstein's convention for summation over repeated indices). 
To compute $\delta\bar\phibf[h]/\delta h$, we take the functional derivative of Eq.~(\ref{phibardef}) wrt $h$, which gives 
\begin{align}
\frac{\delta \bar\phi_i[\x;h]}{\delta h(\y)} ={}& - \int_{\x'}  \Gamma^{(2,0)-1}_{ij}[\x,\x';\bar\phibf[h],h] \nonumber \\ & \times \Gamma_j^{(1,1)}[\x',\y;\bar\phibf[h],h] ,
\label{chi4} 
\end{align} 
where $\Gamma^{(2,0)-1}$ is the propagator. From~(\ref{chi2},\ref{chi3},\ref{chi4}), we finally deduce 
\begin{align}
\chi_s(\y-\y') ={}& - \Gamma^{(0,2)}(\y,\y',\bar\phibf) 
+ \int_{\x,\x'} \Gamma^{(1,1)}_i(\x,\y,\bar\phibf) \nonumber \\ & \times \Gamma^{(2,0)-1}_{ij}(\x,\x',\bar\phibf)  \Gamma^{(1,1)}_j(\x',\y',\bar\phibf) 
\end{align} 
(with $\bar\phibf=\bar\phibf[h=0]$) or, in Fourier space,  
\begin{align}
\chi_s(\p) ={}& -  \Gamma^{(0,2)}(\p,\bar\phibf) \nonumber \\ 
& + \Gamma^{(1,1)}_{i}(\p,\bar\phibf) \Gamma^{(2,0)-1}_{ij}(\p,\bar\phibf)  \Gamma^{(1,1)}_{j}(\p,\bar\phibf) .
\label{chi5} 
\end{align} 
The last term corresponds to the part of the scalar susceptibility which is not 1PI. Equation~(\ref{chi5}) is shown diagrammatically in Fig.~\ref{fig_chis_diag}. 

Since $\phibf$ transforms like a vector under O($N$) rotations while $h$ is a scalar, 
\begin{equation}
\begin{split}
\Gamma_{i}^{(1,1)}(\p,\phibf) &= \phi_i f(\p,\rho) , \\ 
\Gamma^{(0,2)}(\p,\phibf) &= \gamma(\p,\rho) , 
\end{split}
\label{gamdef} 
\end{equation}
where $f$ and $\gamma$ are functions of $p=|\p|$ and the O($N$) invariant $\rho$. To determine the scalar susceptibility in the NPRG approach, we must therefore consider the $k$-dependent vertices $\Gamma_k^{(0,2)}$ and $\Gamma_{k,i}^{(1,1)}$ or, equivalently, the $k$-dependent functions $f_k(\p,\rho)$ and $\gamma_k(\p,\rho)$. 

In the universal regime near the QCP,\cite{Podolsky12} 
\begin{equation}
\begin{split}
\chi_s(\p) &= \calB_\pm + \calA_\pm \Delta^{3-2/\nu} \tilde\Phi_{s,\pm}\left(\frac{p}{\Delta}\right) , \\  
\chi_s''(\w) &= {\rm Im}[ \chi_s^R(\w) ] =   \calA_\pm \Delta^{3-2/\nu} \Phi_{s,\pm}\left(\frac{\w}{\Delta}\right) ,
\end{split}
\label{chis_scaling}
\end{equation}
where $\tilde\Phi_{s,\pm}$ and $\Phi_{s,\pm}$ are universal scaling functions and $\calA_\pm,\calB_\pm$ nonuniversal constants. At the QCP ($\Delta=0$), $\chi_s(\p)-\chi_s(0)\sim p^{3-2/\nu}$ and $\chi''_s(\w)\sim |\w|^{3-2/\nu}$. Since $\chi_s''(\w)$ is an odd function of $\w$, we shall only consider the case $\w>0$ in the following.

\subsection{BMW approximation}

In this section, we first review the BMW approximation for the solution of the RG equation~(\ref{rgeq}) when $h=0$.\cite{Blaizot06,Benitez09,Benitez12} We then show how this approximation can be extended to the calculation of the scalar susceptibility. 

\subsubsection{Effective potential $U_k$ and 2-point vertex $\Gamma_k^{(2,0)}$}

Equation~(\ref{rgeq}) cannot be solved exactly. However, it can be used to derive flow equations for the effective potential and the 1PI vertices. The effective potential is determined by 
\begin{equation}
\dk U_k(\rho) = \half \int_\q \dot R_k(\q) [G_{k,\rm L}(\q,\rho) + (N-1) G_{k,\rm T}(\q,\rho) ] 
\label{eqUk} 
\end{equation}
with initial condition $U_\Lambda(\rho)=r_0\rho+(u_0/6N)\rho^2$. We use the notation 
\beq
\frac{1}{V} \sum_\q \to \int \frac{d^3q}{(2\pi)^3} \equiv \int_\q \quad \mbox{for} \quad V\to\infty. 
\eeq
Solving~(\ref{eqUk}) requires to know the propagator $G_k=(\Gamma_k^{(2,0)}+R_k)^{-1}$ and therefore the 2-point vertex $\Gamma_k^{(2,0)}$ in a uniform field. The latter satisfies the equation (Fig.~\ref{fig_rgeq_diag}) 
\begin{multline}
\dk \Gamma_{k,ij}^{(2,0)}(\p,\phibf) = \sum_{\q}\tdk G_{k,i_1i_2}(\q,\phibf) \\ \times  
\bigg[\half \Gamma^{(4,0)}_{k,iji_2i_1}(\p,-\p,\q,-\q,\phibf) -  \Gamma^{(3,0)}_{k,ii_2i_3}(\p,\q,-\p-\q,\phibf)\\
\times G_{k,i_3i_4}(\p+\q,\phibf) \Gamma^{(3,0)}_{k,ji_4i_1}(-\p,\p+\q,-\q,\phibf)\bigg] ,
\label{rgeq1}
\end{multline}
where 
\begin{equation}
\tdk G_{k,i_1i_2}(\q,\phibf) = - \dot R_k(\q) G_{k,i_1i_3}(\q,\phibf) G_{k,i_3i_2}(\q,\phibf) .
\label{dGk} 
\end{equation}
More generally, the 1PI vertices satisfy an infinite hierarchy of equations since $\dk \Gamma^{(n,0)}_k$ involves  $\Gamma^{(n+1,0)}_k$ and $\Gamma^{(n+2,0)}_k$. The BMW approximation\cite{Blaizot06,Benitez09,Benitez12} allows us to close this infinite hierarchy of equations. It is based on the observation that the $\dot R_k(\q)$ term in~(\ref{dGk}) restricts the integration over the loop momentum in~(\ref{rgeq1}) to small values $|\q|\lesssim k$ whereas the regulator term $\Delta S_k$ ensures that the vertices $\Gamma_k^{(n,0)}(\p_1\cdots\p_n,\phibf)$ are regular functions of the momenta when $|\p_i|\ll k$. This enables us to set $\q=0$ in the vertices $\Gamma_k^{(3,0)}$ and $\Gamma_k^{(4,0)}$ in~(\ref{rgeq1}). Noting that in a uniform field, 
\begin{equation}
\begin{split}
\Gamma_{k,i_1i_2i_3}^{(3,0)}(\p,-\p,0,\phibf) &= \frac{1}{\sqrt{V}} \frac{\partial \Gamma^{(2,0)}_{k,i_1i_2}(\p,-\p,\phibf)}{\partial \phi_{i_3}} , \\  
\Gamma_{k,i_1i_2i_3i_4}^{(4,0)}(\p,-\p,0,0,\phibf) & =\frac{1}{V} \frac{\partial^2 \Gamma^{(2,0)}_{k,i_1i_2}(\p,-\p,\phibf)}{\partial\phi_{i_3} \partial\phi_{i_4}} , 
\end{split}
\end{equation}
we obtain a closed equation for the 2-point vertex, 
\begin{multline}
\dk \Gamma_{k,ij}^{(2,0)}(\p,\phibf) = \half \int_\q [\tdk G_{k,i_1i_2}(\q,\phibf)]  \frac{\partial^2 \Gamma_{k,ij}^{(2,0)}(\p,\phibf)}{\partial\phi_{i_1}\partial\phi_{i_2}} \\
- \int_\q [\tdt G_{k,i_1i_2}(\q,\phibf)] \frac{\partial \Gamma_{k,ii_3}^{(2,0)}(\p,\phibf)}{\partial\phi_{i_2}} \\
 \times G_{k,i_3i_4}(\p+\q,\phibf) \frac{\partial \Gamma_{k,ji_4}^{(2,0)}(\p,\phibf)}{\partial\phi_{i_1}} ,
\label{rgeq2}
\end{multline}
which must be solved together with~(\ref{eqUk}). 

Equation~(\ref{rgeq2}) yields two coupled equations for $\GamA(\p,\rho)$ and $\GamB(\p,\rho)$ [see Eq.~(\ref{GamAB})]. Following Refs.~\onlinecite{Blaizot06,Benitez09,Benitez12}, we introduce self-energies $\Delta_{A,k}$ and $\Delta_{B,k}$ defined by 
\begin{equation}
\begin{split} 
\GamA(\p,\rho) &= \p^2 + \DA(\p,\rho) + U_k'(\rho) , \\
\GamB(\p,\rho) &= \DB(\p,\rho) + U_k''(\rho) ,
\end{split}
\end{equation}
with $\DA(\p=0,\rho)=\DB(\p=0,\rho)=0$. The effective potential $U_k$ is obtained from the (exact) RG equation~(\ref{eqUk}) while the self-energies are deduced from the (approximate) RG equation~(\ref{rgeq2}). The final form of the equations is obtained by writing the self-energies as\cite{Blaizot06,Benitez09,Benitez12}
\begin{equation}
\begin{split}
\DA(\p,\rho) &= \p^2 \YA(\p,\rho)  , \\ 
\DB(\p,\rho) &= \p^2 \YB(\p,\rho) ,
\end{split} 
\end{equation}
and solve for $\YA$ and $\YB$ with initial conditions $Y_{A,\Lambda}(\p,\rho)=Y_{B,\Lambda}(\p,\rho)=0$.

\subsubsection{Scalar susceptibility} 
\label{subsec_chis} 

\begin{figure}
\centerline{\includegraphics[width=6cm]{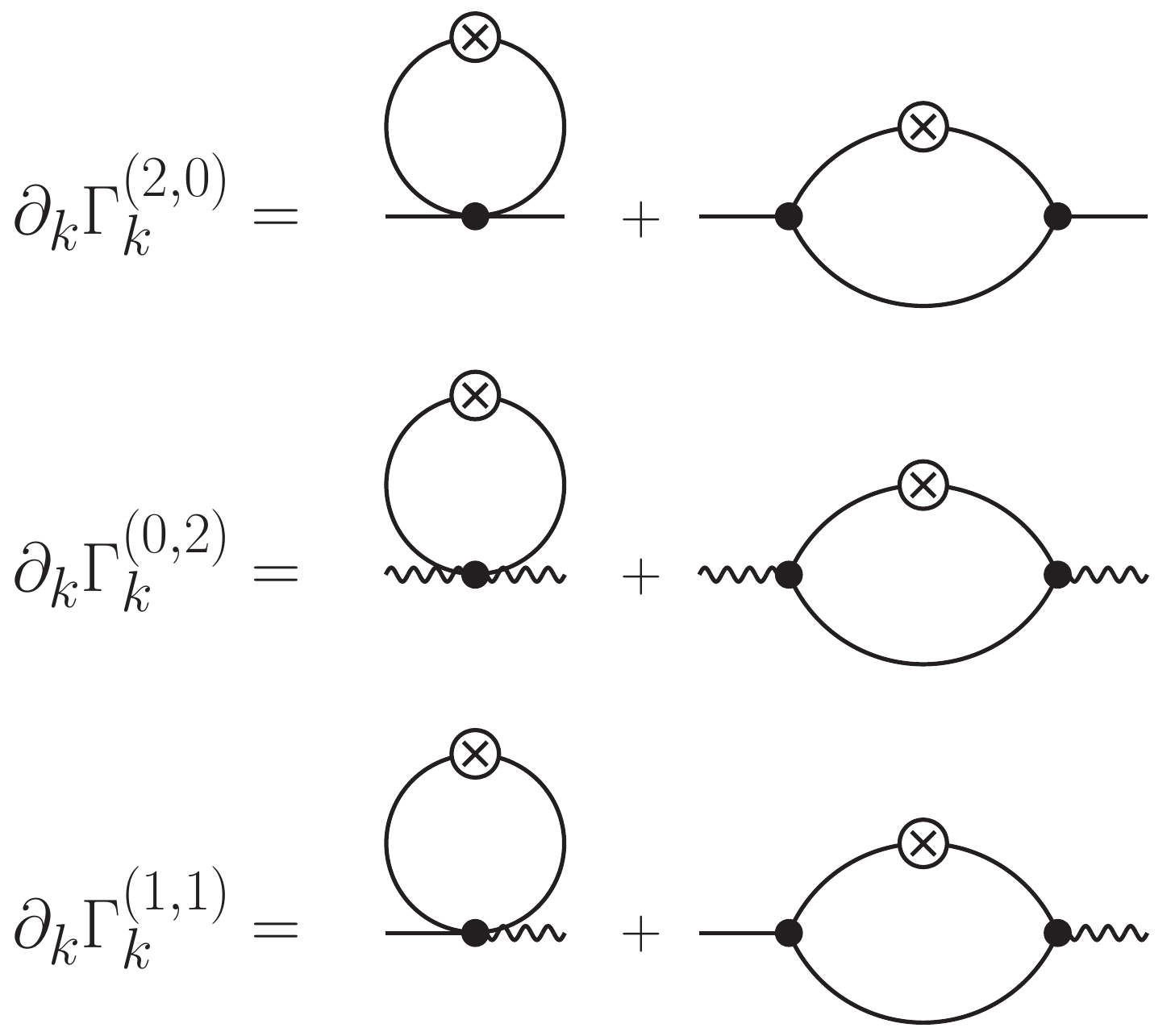}}
\caption{Diagrammatic representation of the RG equations~(\ref{rgeq1},\ref{rgeq5},\ref{rgeq6}). Signs and symmetry factors are not shown. The diagrammatic representation of the vertices $\Gamma_k^{(n,m)}$ is the same as in Fig.~\ref{fig_chis_diag} and the cross stands for $\dk R_k$.}
\label{fig_rgeq_diag} 
\end{figure}

The vertices $\Gamma_k^{(0,2)}$ and $\Gamma_k^{(1,1)}$ in a constant field satisfy the RG equations 
\begin{multline}
\dk \Gamma_{k}^{(0,2)}(\p,\phibf) = \sum_{\q}\tdk G_{k,i_1i_2}(\q,\phibf) \\ \times  
\bigg[\half \Gamma^{(2,2)}_{k,i_2i_1}(\q,-\q,\p,-\p,\phibf) -  \Gamma^{(2,1)}_{k,i_2i_3}(\q,-\p-\q,\p,\phibf)\\
\times G_{k,i_3i_4}(\p+\q,\phibf) \Gamma^{(2,1)}_{k,i_4i_1}(\p+\q,-\q,-\p,\phibf)\bigg] ,
\label{rgeq3}
\end{multline}
and 
\begin{multline}
\dk \Gamma_{k,i}^{(1,1)}(\p,\phibf) = \sum_{\q}\tdk G_{k,i_1i_2}(\q,\phibf) \\ \times  
\bigg[\half \Gamma^{(3,1)}_{k,ii_2i_1}(\p,\q,-\q,-\p,\phibf) -  \Gamma^{(3,0)}_{k,i_2i_3}(\p,\q,-\p-\q,\phibf)\\
\times G_{k,i_3i_4}(\p+\q,\phibf) \Gamma^{(2,1)}_{k,i_4i_1}(\p+\q,-\q,-\p,\phibf)\bigg] .
\label{rgeq4}
\end{multline}
The BMW approximation amounts to setting $\q=0$ in the vertices $\Gamma_k^{(n,m)}$ in~(\ref{rgeq3}) and (\ref{rgeq4}). Using 
\begin{equation}
\begin{split}
\Gamma_{k,i_1i_2}^{(2,1)}(\p,0,-\p,\phibf) &= \frac{1}{\sqrt{V}} \frac{\partial \Gamma^{(1,1)}_{k,i_1}(\p,-\p,\phibf)}{\partial \phi_{i_2}} , \\ 
 \Gamma_{k,i_1i_2}^{(2,2)}(0,0,\p,-\p,\phibf) & =\frac{1}{V} \frac{\partial^2 \Gamma^{(0,2)}_{k}(\p,-\p,\phibf)}{\partial\phi_{i_1} \partial\phi_{i_2}} , \\  
\Gamma_{k,i_1i_2i_3}^{(3,1)}(\p,0,0,-\p,\phibf) & =\frac{1}{V} \frac{\partial^2 \Gamma^{(1,1)}_{k,i_1}(\p,-\p,\phibf)}{\partial\phi_{i_2} \partial\phi_{i_3}} , 
\end{split}
\end{equation}
one then obtains closed equations for $\Gamma_k^{(0,2)}$ and $\Gamma_k^{(1,1)}$:
 \begin{multline}
\dk \Gamma_{k}^{(0,2)}(\p,\phibf) = \half \int_{\q}\tdk G_{k,i_1i_2}(\q,\phibf) \frac{\partial^2\Gamma^{(0,2)}_{k}(\p,\phibf)}{\partial\phi_{i_2}\partial\phi_{i_1}}  \\ 
- \int_\q \tdk G_{k,i_1i_2}(\q,\phibf) \frac{\partial\Gamma^{(1,1)}_{k,i_3}(\p,\phibf)}{\partial\phi_{i_2}} \\ 
\times G_{k,i_3i_4}(\p+\q,\phibf) \frac{\partial\Gamma^{(1,1)}_{k,i_4}(\p,\phibf)}{\partial\phi_{i_1}} ,
\label{rgeq5}
\end{multline}
and 
\begin{multline}
\dk \Gamma_{k,i}^{(1,1)}(\p,\phibf) = \half \int_{\q}\tdk G_{k,i_1i_2}(\q,\phibf)  
\frac{\partial^2 \Gamma^{(1,1)}_{k,i}(\p,\phibf)}{\partial\phi_{i_2}\partial\phi_{i_1}} \\ 
- \int_\q \tdk G_{k,i_1i_2}(\q,\phibf)  \frac{\partial \Gamma^{(2,0)}_{k,ii_3}(\p,\phibf)}{\partial \phi_{i_2}} \\ 
\times G_{k,i_3i_4}(\p+\q,\phibf) \frac{\partial\Gamma^{(1,1)}_{k,i_4}(\p,\phibf)}{\partial\phi_{i_1}} .
\label{rgeq6}
\end{multline}
Equations~(\ref{rgeq5}) and (\ref{rgeq6}) are shown in Fig.~\ref{fig_rgeq_diag}. They lead to RG equations for the functions $f_k$ and $\gamma_k$ defined in~(\ref{gamdef}) with initial conditions $f_\Lambda(\p,\rho)=-2$ and $\gamma_\Lambda(\p,\rho)=0$. 

\subsection{Explicit form of RG equations} 
\label{subsec_explicit_rg} 

In this section, we give the explicit forms of the flow equations in the BMW approximation. We define the threshold functions
\begin{align} 
J_{k,n}^{\alpha\beta}(\p,\rho) &= \int_\q [\dt R_k(\q)] G_{k,\alpha}^{n-1}(\q,\rho) G_{k,\beta}(\p+\q,\rho) , \nonumber \\ 
I_{k,n}^{\alpha\beta}(\rho) &=J_{k,n}^{\alpha\beta}(\p=0,\rho) , 
\label{threshold}
\end{align} 
where $\alpha,\beta={\rm L,T}$ and $t=\ln(k/\Lambda)$ is a (negative) RG ``time". 

The derivative $W_k(\rho)=U_k'(\rho)$ of the effective potential satisfies the equation  
\begin{equation}
\dt W = \half \bigl[ I^{\rm LL}_1{}' + (N-1) I^{\rm TT}_1{}' \bigr] ,
\label{rgeq7}
\end{equation}
where the primes denote a $\rho$ derivative. Here and in the following, to alleviate the notations, we do not write the $k$, $\p$ and $\rho$ dependence of the functions  (and use the notation $p=|\p|$). The 2-point vertex is determined by the flow equations 
\begin{align} 
\partial_t Y_{\rm A} ={}& \dfrac{2 \rho}{p^2}\bigl[
J_3^{\rm LT}(p^2 Y_{\rm A}' +W')^2 + J_3^{\rm TL}(p^2 Y_{\rm B} +W')^2 \nonumber \\ & - (I_3^{\rm LT}+I_3^{\rm TL})W'^2
\bigr] -\dfrac{1}{2}I_2^{\rm LL}(Y_{\rm A}'+2\rho Y_{\rm A}'') \nonumber \\
& -\dfrac{1}{2}I_2^{\rm TT}[(N-1)Y_{\rm A}'+2Y_{\rm B}] 
\label{rgeq8}
\end{align} 
and 
\begin{align} 
\partial_t Y_{\rm B} ={}& \dfrac{1}{p^2}\bigl\{(N-1)\bigl[J_3^{\rm TT}(p^2 Y_{\rm B} +W')^2-I_3^{\rm TT}W'^2 \bigr] \nonumber \\
& + J_3^{\rm LL}[p^2(Y_{\rm A}'+2Y_{\rm B}+2\rho Y_{\rm B}')+3W'+2\rho W'']^2 \nonumber \\ & 
-I_3^{\rm LL}(3W'+2\rho W'')^2 - J_3^{\rm LT}(p^2 Y_{\rm A}' +W')^2  \nonumber \\ & 
+ I_3^{\rm LT}W'^2 - J_3^{\rm TL}(p^2 Y_{\rm B} +W')^2 + I_3^{\rm TL}W'^2\bigr\} \nonumber\\
& -\dfrac{1}{2}I_2^{\rm TT}(N-1)Y_{\rm B}'-\dfrac{1}{2}I_2^{\rm LL}(5Y_{\rm B}'+2\rho Y_{\rm B}'') \nonumber \\ & 
+ \frac{1}{2\rho} ( I_{2}^{\rm TT} - I_{2}^{\rm LL} ) Y_{\rm B} .
\label{rgeq9}
\end{align} 
Note that the rhs of the last equation is well defined for $\rho\to 0$ and $p\to 0$.\cite{Benitez12}
Equations~(\ref{rgeq7}-\ref{rgeq9}) were previously derived in Refs.~\onlinecite{Blaizot06,Benitez09,Benitez12}.

$f$ and $\gamma$ satisfy the equations 
\begin{align} 
\partial_t f ={}&  J_3^{\rm LL} (f+2\rho f')[p^2(Y_{\rm A}'+2Y_{\rm B}+2\rho Y_{\rm B}')+3W' \nonumber \\ & 
+2\rho W''] + (N-1)J_3^{\rm TT} f (p^2Y_{\rm B} + W')- I_2^{\rm LL} \rho f'' \nonumber \\ &  
-\dfrac{1}{2} f'[3 I_2^{\rm LL} +(N-1) I_2^{\rm TT}]
\end{align} 
and 
\begin{align} 
\partial_t \gamma ={}& (N-1)J_3^{\rm TT} f^2 + J_3^{\rm LL}(f+2\rho f')^2  \nonumber \\ &
-\dfrac{1}{2} [I_2^{\rm LL}(\gamma'+2\rho\gamma'')+(N-1)I_2^{\rm TT}\gamma'].
\end{align}

The QCP manifests itself as a fixed point of the RG equations provided we use dimensionless flow equations where all quantities are expressed in units of the running scale $k$. We therefore use the following dimensionless variables (with $d=3$) 
\begin{equation}
\tilde p=k^{-1} p, \qquad \trho = K_d^{-1} Z_k k^{2-d} \rho , 
\end{equation}
and functions\cite{not4} 
\begin{equation}
\begin{gathered} 
\tilde U_k(\trho) =  K_d^{-1}k^{-d} U_k(\rho), \\ 
\tilde W_k(\trho) =  Z_k^{-1} k^{-2} W_k(\rho) , \\ 
1+\tilde Y_{A,k}(\tilde p,\trho) = Z_k^{-1} [1+Y_{A,k}(p,\rho)] , \\
\tilde Y_{B,k}(\tilde p,\trho) = K_d Z_k^{-2} k^{d-2} Y_{B,k}(p,\rho) , \\
\tilde f_k(\tilde p,\trho) = f_k(p,\rho), \\ 
\tilde\gam_k(\tilde p,\trho) =  K_d^{-1} Z_k^2 k^{4-d} \gam_k(p,\rho) , 
\end{gathered}
\label{vdim}
\end{equation}  
where $K_d^{-1}=2^{d-1} d\pi^{d/2}\Gamma(d/2)$ is a constant originating from angular integrals which is introduced here for convenience. The field renormalization factor $Z_k$ is defined by 
\begin{equation}
Z_k = \frac{\partial \Gamma^{(2,0)}_{k,\rm T}(\p,\rho)}{\partial p^2}\biggl|_{\p=0, \rho=\rho_{0,k}} 
= 1 +Y_{A,k}(\p=0,\rho_{0,k}) ,
\label{Zkdef}
\end{equation}
where $\rho_{0,k}$ corresponds to the minimum of the effective potential $U_k(\rho)$. Equation~(\ref{Zkdef}) implies that
\begin{equation}
\Gamma^{(2,0)}_{k,\rm T}(\p,\rho_{0,k})\simeq Z_k\p^2+W_k(\rho_{0,k}) \quad \mbox{for} \quad \p\to 0
\label{gamT}
\end{equation}
so that in the ordered phase the stiffness is simply defined by $\rho_{s,k}=2Z_k\rho_{0,k}$ [Eq.~(\ref{rhosdef})]. If we use~(\ref{gamT}) to estimate the gap $\Delta$ in the disordered phase (where $\rho_{0,k=0}=0$), we find 
\begin{equation}
\Delta = \left( \frac{W_{k=0}(0)}{Z_{k=0}} \right)^{1/2} . 
\label{Deltaest} 
\end{equation}
The numerical solution of the flow equations show that this expression is in very good agreement (with an error smaller than 1\textperthousand) with the exact determination of the gap obtained from the peak in the spectral function $\chi''_{\rm L,T}(\w)$ (Sec.~\ref{subsubsec_disordered}). 

We also define a (running) anomalous dimension
\begin{equation}
\eta_k = - k \dk \ln Z_k . 
\end{equation}
At criticality ($r_0=r_{0c}$), the anomalous dimension is obtained from $\eta=\lim_{k\to 0} \eta_k$. The normalization condition~(\ref{Zkdef}) can be expressed as $\tilde Y_{A,k}(\tilde p=0,\trho_{0,k})=0$ which provides us with an equation for $\eta_k$. This equation takes a simpler form if we choose $\rho=0$ in the definition~(\ref{Zkdef}) of $Z_k$.\cite{Benitez12} This choice is however not appropriate in the ordered phase where we are eventually interested in quantities defined at the nonzero order parameter $\rho_{0,k}$, which corresponds to a diverging dimensionless field $\trho_{0,k}\sim \rho_{0,k}/k$.

\section{Large-$N$ limit} 
\label{sec_largeN} 

In this section, following Ref.~\onlinecite{Blaizot06}, we show that the BMW equations become exact and can be solved analytically when $N\to\infty$. 
In this limit, $\Gamma_k[\phibf]$ is of order $N$ and the field $\phibf$ is of order $\sqrt{N}$ (see Appendix~\ref{app_largeN}). This implies that $\gamma_k$ is $\calO(N)$, $W_k$, $\GamA$ and $f_k$ are $\calO(1)$ (as well as the threshold functions $I_{k,n}^{\alpha\beta}$ and $J_{k,n}^{\alpha\beta}$) whereas 
 $\GamB$ is $\calO(1/N)$. It follows that 
\begin{equation}
\dt Y_{A,k} = - \frac{N}{2} I_2^{\rm TT} Y_{A,k}' .
\end{equation}
Since $Y_{A,\Lambda}=0$, we deduce $Y_{A,k}=\Delta_{A,k}=0$: the momentum dependence of the transverse propagator is not renormalized in the large-$N$ limit. The other equations read
\begin{equation}
\begin{split} 
\dt W_k &= \frac{N}{2} I_1^{\rm TT}{}' , \\ 
\dt \GamB &= N J_3^{\rm TT} \GamB^2 - \frac{N}{2}I_2^{\rm TT} \GamB' , \\ 
\dt f_k &=  N J_3^{\rm TT} f_k \GamB - \frac{N}{2} I_2^{\rm TT} f_k' , \\
\dt \gam_k &= N J_3^{\rm TT} f_k^2 - \frac{N}{2} I_2^{\rm TT} \gam_k' .
\end{split}
\end{equation}
To obtain the equation for $\GamB$, we have used $I_1^{\rm TT}{}'=-W_k' I_2^{\rm TT}$ and $I_2^{\rm TT}{}'=-2W_k' I_3^{\rm TT}$ when $\DA=0$. To solve these equations, we set $W=W_k(\rho)$ and use the variables $(k,W)$ instead of $(k,\rho)$.\cite{Blaizot06} This is done introducing the function $g_k(W)=\rho$ and using $g_k'(W)=1/W_k'(\rho)$, 
\begin{align}
\dt g_k(W) &= - \frac{1}{W_k'(\rho)} \dt W_k(\rho) = \frac{N}{2} I_2^{\rm TT} \nonumber \\ 
&=  \frac{N}{2} \int_\q \frac{\dt R_k(\q)}{[\q^2 + W +R_k(\q)]^2} .
\end{align}
Since $k$ and $W$ are independent variables, this equation can be rewritten as 
\begin{equation}
\dt g_k(W) = - \frac{N}{2} \dt \int_\q \frac{1}{\q^2 + W +R_k(\q)} , 
\label{largeN1}
\end{equation}
where both sides are total derivatives, and we obtain
\begin{align}
g_k(W) - g_\Lambda(W) ={}& - \frac{N}{2}\int_\q \biggl( \frac{1}{\q^2 + W +R_k(\q)} \nonumber \\ & 
- \frac{1}{\q^2 + W +R_\Lamb(\q)} \biggr) .
\label{largeN2}
\end{align} 
For $R_\Lamb(\q)\to\infty$, using $g_\Lambda(W)=(3N/u_0)(W-r_0)$ we reproduce the known result in the large-$N$ limit [Eq.~(\ref{app4})]. For $R_\Lamb(\q)<\infty$, we obtain an apparent difference with the exact result, which is explained by the fact that $\Gamma_\Lambda[\phibf]$  is not given by $S[\phibf]$. This does not matter when one is interested in universal properties in the vicinity of the QCP: the microscopic physics can be directly parameterized by $\Gamma_\Lambda[\phibf]$ (which we can choose to coincide with $S[\phibf]$). In that case, however, it is important to keep the term with $R_\Lamb$ in~(\ref{largeN2}) for comparison with the numerical solution of the flow equations in the large-$N$ limit.

To calculate $\GamB$, we use 
\begin{equation}
\dt \GamB\bigl|_W = \dt \GamB\bigl|_\rho + \GamB' \dt g_k = N J_3^{\rm TT} \GamB^2 
\end{equation}
(the prime denotes a $\rho$ derivative), i.e. 
\begin{equation}
\dt \GamB^{-1} \bigl|_W =  -N J_3^{\rm TT} . 
\end{equation}
The rhs is again a total derivative. Using $\GamBL=u_0/3N$, we obtain 
\begin{equation}
\GamB(\p,\rho) = \frac{1}{\frac{3N}{u_0} + \frac{N}{2} [ \Pi_k(\p,\rho) - \Pi_\Lambda(\p,\rho) ] }
\label{largeN3}
\end{equation}
where 
\begin{align}
\Pi_k(\p,\rho) ={}& \int_\q \frac{1}{\q^2+W_k(\rho)+R_k(\q)} \nonumber \\ &
\times \frac{1}{(\p+\q)^2+W_k(\rho)+R_k(\p+\q)} .
\end{align} 
We recover the large-$N$ result derived in Appendix~\ref{app_largeN} when $R_\Lamb(\q)\to\infty$ and $\Pi_\Lambda\to 0$.

A similar procedure is used to calculate $f_k$ and $\gam_k$. From 
\begin{equation}
\begin{split} 
\dt f_k \bigl|_W &= N J_3^{\rm TT} \GamB f_k , \\ 
\dt \gam_k \bigl|_W &=N J_3^{\rm TT} f_k^2 ,
\end{split} 
\end{equation} 
and the initial conditions $f_\Lambda=-2$, $\gam_\Lamb=0$, we obtain 
\begin{equation}
\begin{split} 
f_k(\p,\rho) &= -\frac{6N}{u_0} \GamB(\p,\rho) \\
\gam_k(\p,\rho) &= -\frac{12N}{u_0} + \left(\frac{6N}{u_0}\right)^2 \GamB(\p,\rho) , 
\end{split}
\label{largeN4}
\end{equation} 
in agreement (when $R_\Lambda(\q)\to\infty$) with the results~(\ref{app5},\ref{app6}) obtained in the standard large-$N$ approach. 

The numerical solution of the flow equations in the large-$N$ limit is discussed in Sec.~\ref{subsec_num_largeN}.

\section{Higgs and longitudinal spectral functions} 
\label{sec_spectral} 

We solve numerically the flow equations with $u_0=200$ and $\Lambda=1$. We use a $(\tilde p,\trho)$ grid of $80\times 200$ points with $0\leq \tilde p\leq \tilde p_{\rm max}$, $0\leq \trho\leq \trho_{\rm max}$, $\tilde p_{\rm max}=8$ and $\trho_{\rm max}=6N$. The flow equations are integrated using an explicit Euler method with $\Delta t=-10^{-4}$ ($t=\ln(k/\Lambda)$). We use Simpson's rule to compute momentum integrals (with an upper cutoff $\tilde q_{\rm max}=4$) and finite-difference evaluation for derivatives. We have verified the stability of our results wrt to the various parameters used (number of points in the $(\tilde p,\trho)$ grid, upper cutoff in momentum integrals, etc.).

The ordered phase requires some care. First, since $\rho_{0,k}$ converges towards a nonzero value $\rho_0$ for $k\to 0$, $\trho_{0,k}\sim \rho_0/k$ diverges and one cannot work with a fixed $\trho$ grid. To circumvent this difficulty, we start the flow with a fixed $\trho$ grid but switch to a fixed $\rho$ grid ($0\leq \rho\leq\rho_{\rm max}$) once the flow leaves the critical regime to reach the ordered regime, which occurs for $k$ of the order of the inverse of the Josephson length $\xi_J\sim(r_{0c}-r_0)^{-\nu}$. Second, since $W_k(\rho_{0,k})=0$, $W_k(\rho)$ is negative for $\rho<\rho_{0,k}$ and $G_k(\p=0,\rho=0)=(R_k(\p=0)+W_k(0))^{-1}$ becomes very large. This behavior is associated to the approach to convexity of the 
effective potential $U_k(\rho)$.\cite{Berges02} While $G_k(\p,\rho)$ should remain strictly positive, we find that it develops a pole for small $k$ in the BMW approximation, which leads to a divergence of the threshold functions~(\ref{threshold}) and a numerical instability.\cite{[{Note that this problem is not specific to the BMW approximation. For example, it also appears in the NPRG analysis of the two-dimensional O(2) model based on a derivative expansion of the scale-dependent effective action, see }] Jakubczyk14} 
To cure this problem, we eliminate from the $\rho$ grid the points for which $W_k(\rho)$ is smaller than $-2$.\cite{not7} The grid $[\rho_{{\rm min},k},\rho_{\rm max}]$ becomes $k$ dependent.\cite{[{A $k$-dependent grid has also been used by G. v. Gersdorff in order to avoid numerical instabilities in the NPRG study of the two-dimensional O(2) model: }] Gersdorff00,*Gersdorff01}
When $\rho_{{\rm min},k}$ becomes nearly equal to $\rho_{0,k}$, the flow cannot be continued anymore since the minimum of the effective potential must remain in the range $[\rho_{{\rm min},k},\rho_{\rm max}]$ for physical quantities (defined for $\rho=\rho_{0,k}$) to be determined. However, this procedure allows us to reach small values of $k$ (typically $k_{\rm min}\simeq 0.05\Delta$) for which physical quantities have nearly converged to their $k=0$ values (the only exception is the very low-energy behavior $p,|\w|\ll\Delta$ of correlation functions, see below).\cite{not6} A more precise estimate of the $k=0$ values can be obtained using an extrapolation of the type $a+bk^c$.

A momentum dependent function $F(p)\equiv F_{k=0}(p)$, such as a two-point vertex or the scalar susceptibility, is obtained from the approximation $F(p)\simeq F_{k=p/\tilde p_{\rm max}}(p)$ where $k=p/\tilde p_{\rm max}$ is the smallest value of $k$ for which the dimensionless momentum $\tilde p=p/k$ is still in the grid $[0,\tilde p_{\rm max}]$. We have verified, by increasing $\tilde p_{\rm max}$, that the flow of $F_k(p)$ for  $k<p/\tilde p_{\rm max}$ is negligible. This is due to the fact that, in the cases we are considering here, $p$ acts as an effective infrared cutoff, while only momenta of the order of $k$ or smaller contribute to the flow, so that the flow of the function $F_k(p)$ effectively stops when $k\ll p$. To obtain the retarded dynamical function $F^R(\w)$, we compute $F(p)$ for $M$ momentum values $p_l$ ($l=1\cdots M$) with typically $M$ in the range $50-100$. One 
then constructs a $M$-point Pad\'e approximant $F_P(p)$ which coincides with $F(p)$ for all $p_l$'s, and $F^R(\w)$ is approximated by $F_P(-i\w+\eps)$ (we take $\eps/\Delta<10^{-4}$).\cite{Vidberg77} Note that in the ordered phase, we cannot determine $F(p)$ for values of $p$ below $k_{\rm min}\tilde p_{\rm max}=8k_{\rm min}\simeq 0.4\Delta$. This prevents us to determine $F^R(\w)$ for $\w\lesssim 0.4\Delta$. On the other hand, in the disordered phase, the spectral functions we are interested in vanish for $\w<\Delta$ or $\w<2\Delta$ and there is no need to compute $F(p)$ and $F^R(\w)$ for $p$ or $\w$ smaller than $0.4\Delta$. 

\begin{figure}
\centerline{\includegraphics[width=7.8cm]{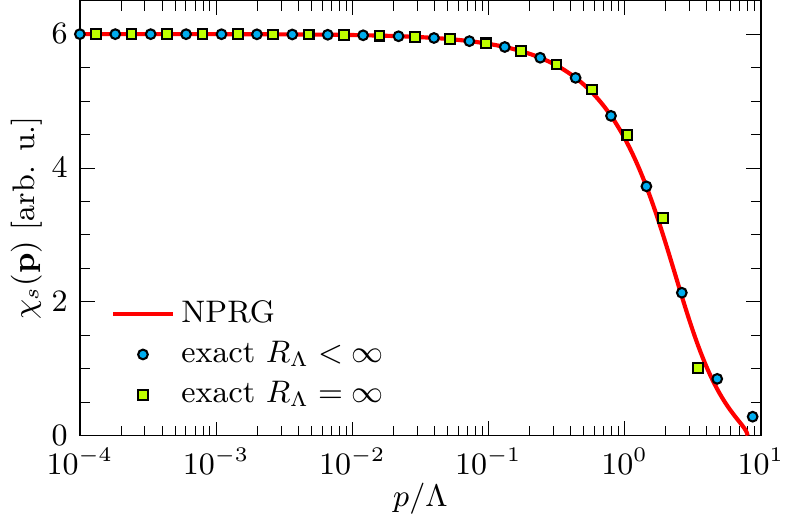}}
\caption{(Color online) Scalar susceptibility $\chi_s(\p)$ at the QCP for $N=1000$ (solid line) compared to the exact large-$N$ solution with $R_\Lambda<\infty$ (circles) and $R_\Lambda=\infty$ (squares). Here and in the following figures we use arbitrary units for the scalar and longitudinal susceptibilities.}
\label{fig_Ngrd_chi_QCP}
\end{figure}

\subsection{Large-$N$ limit} 
\label{subsec_num_largeN}

As a check of our procedure we first discuss the numerical solution of the flow equations in the large-$N$ limit where comparison with exact results is possible (Sec.~\ref{sec_largeN}). Figure~\ref{fig_Ngrd_chi_QCP} shows the scalar susceptibility $\chi_s(\p)$ obtained for $N=1000$ at the QCP. Except for momenta near the cutoff $\Lambda$, we obtain a very good agreement with the exact solution~(\ref{largeN4}) in the limit $N\to\infty$ taking into account the finite value of $R_\Lambda(\p)$. For sufficiently small $p$, when $\Pi_{k=0}(\p)$ becomes very large, the scalar susceptibility becomes independent of the initial value $R_\Lambda$ of the cutoff function. In any case, for universal properties, the value of $R_\Lambda$ does not matter. 

\begin{figure}
\centerline{\includegraphics{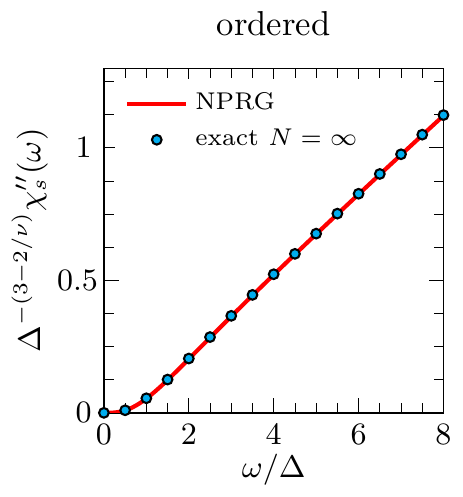}
\includegraphics{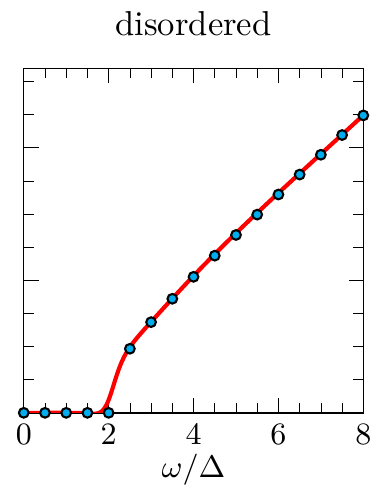}}
\vspace{0.25cm}
\centerline{\includegraphics{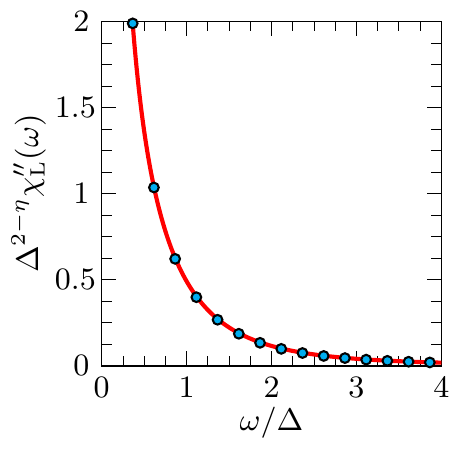}
\includegraphics{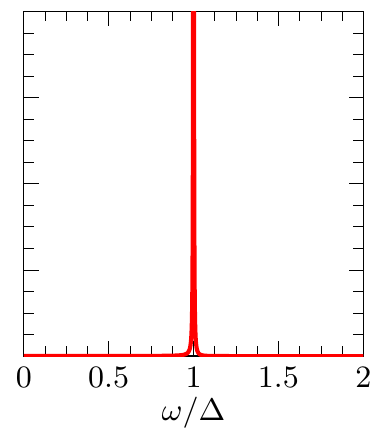}}
\caption{(Color online) Spectral functions $\chi''_s(\omega)$ and $\chi_{\rm L}''(\omega)$ in the ordered and disordered phases for $N=1000$ (solid line), compared to the exact large-$N$ solution (symbols). In the disordered phase, the exact solution for $\chi_{\rm L}''(\omega)=\chi_{\rm T}''(\omega) \sim \delta(\omega-\Delta)$ is not shown. }
\label{fig_Ngrd_chi}
\end{figure}

The spectral functions $\chi''_s(\omega)$ and $\chi_{\rm L}''(\omega)$ are shown in Fig.~\ref{fig_Ngrd_chi} for $N=1000$, in both the ordered and disordered phases, in the universal regime near the QCP. Again, the agreement with the exact results (including nonuniversal prefactors) in the limit $N\to\infty$ is very good. This validates our procedure to compute the momentum dependence of correlation functions as well as the Pad\'e method to obtain the spectral functions. 

\subsection{$N<\infty$}

In the following, we discuss the NPRG results obtained for finite $N$, in particular $N=2$ and $N=3$. 

\subsubsection{QCP}
\label{subsubsec_qcp}

We first solve the equations to determine $r_{0c}$ and the critical exponents $\nu$ and $\eta$. The anomalous dimension $\eta$ is directly obtained from $\lim_{k\to 0}\eta_k$ when $r_0=r_{0c}$. The correlation-length exponent $\nu$ is deduced from the behavior of $\tilde W_k(0)\simeq \tilde W^*(0)+ C e^{-t/\nu}$ at very long time $|t|$ (since the condition $r_0=r_{0c}$ is never exactly fulfilled the RG trajectories will always eventually flow away from the fixed point with an escape rate given by $1/\nu$). Our results agree both with previous NPRG-BMW calculations\cite{Benitez12} and Monte Carlo estimates (see Tables~\ref{table_nu} and \ref{table_eta}). 

\begin{table}
\centering
\setlength{\tabcolsep}{8pt}
\caption{Critical exponent $\nu$ obtained in the NPRG approach, from either $W_k(0)$ (Sec.~\ref{subsubsec_qcp}) or $\chi_s$ (Sec.~\ref{subsubsec_ordered}), compared to Monte Carlo (MC) simulations.}
\begin{tabular}{cccc}
\hline \hline
$N$ & from $W_k(0)$ & from $\chi_s$ & MC \\
\hline
2 & 0.673 & 0.674 & 0.6717(1)~[\onlinecite{Campostrini06}] \\
3 & 0.714 & 0.722 & 0.7112(5)~[\onlinecite{Campostrini02}]\\
4 & 0.754 & 0.766 & 0.749(2)~[\onlinecite{Hasenbusch01}]\\
5 & 0.787 & 0.804 & \\
6 & 0.816 & 0.835 & \\
8 & 0.860 & 0.879& \\
10 & 0.893 & 0.906& \\
100 & 0.990	& 0.992& \\
1000 & 0.999 & 0.999 & \\
\hline \hline
\end{tabular}
\label{table_nu}
\caption{Same as Table~\ref{table_nu} but for the anomalous dimension $\eta$ (defined by $\eta=\lim_{k\to 0}\eta_k$ when $r_0=r_{0c}$).}
\begin{tabular}{ccc}
\hline \hline
$N$ & NPRG & MC\\
\hline
2 & 0.0423 & 0.0381(2)~[\onlinecite{Campostrini06}] \\
3 & 0.0411 & 0.0375(5)~[\onlinecite{Campostrini02}] \\
4 & 0.0386 & 0.0365(10)~[\onlinecite{Hasenbusch01}] \\
5 & 0.0354 &    \\
6 & 0.0321 &    \\
8 & 0.0264 &    \\
10 & 0.0220 &    \\
100 & 0.00233 &    \\
1000 & 0.000233 &    \\
\hline \hline
\end{tabular}
\label{table_eta}
\end{table}

\begin{figure}
\centerline{\includegraphics{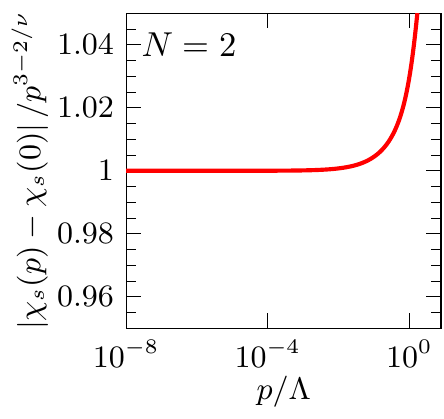}
\includegraphics{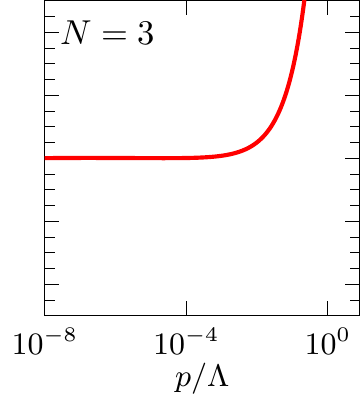}}
\vspace{0.25cm}
\centerline{\includegraphics{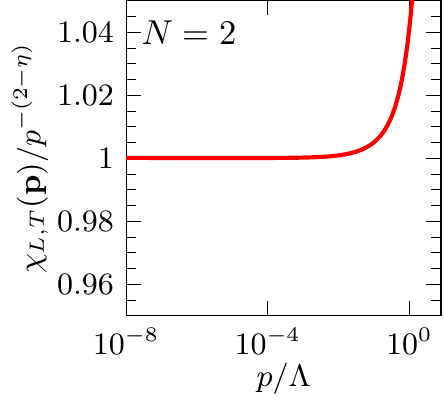}
\includegraphics{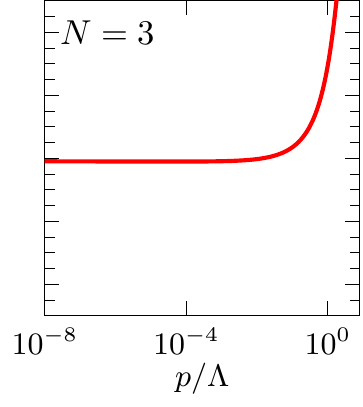}}
\caption{(Color online) $|\chi_s(\p)-\chi_s(0)|/p^{3-2/\nu}$ and $\chi_{\rm L,T}(\p)/p^{-2+\eta}$ at the QCP for $N=2$ and $N=3$. The normalization is chosen to have a ratio equal to one for $p\to 0$.}
\label{fig_chi_qcp}
\end{figure}

Figure~\ref{fig_chi_qcp} shows $\chi_{\rm L,T}(\p)$ and $\chi_s(\p)$ at criticality for $N=2$ and $N=3$. In the universal regime $p\ll p_G$, where $p_G$ is the inverse of the Ginzburg length $\xi_G\sim 24\pi/u_0$, we find $\chi_{\rm L,T}(\p) \sim 1/p^{2-\eta}$ where the value of the anomalous dimension, $\eta\simeq 0.0423$ for $N=2$ and $\eta\simeq 0.0411$ for $N=3$, agrees with the estimate obtained from the running anomalous dimension $\eta_k$ (Table~\ref{table_eta}). As for the scalar susceptibility, we find $\chi_s(\p)\sim p^\theta$ with $\theta\simeq 0.0345$ for $N=2$ and $\theta\simeq 0.230$ for $N=3$. If we use the expected relation $\theta=3-2/\nu$ [Eqs.~(\ref{chis_scaling})], we obtain $\nu\simeq 0.674$ for $N=2$ and $\nu\simeq 0.722$ for $N=3$, in very good agreement with our previous estimates of $\nu$ based on the behavior of $\tilde W_k(0)$ in the close vicinity of the fixed point (Table~\ref{table_nu}).

\subsubsection{Disordered phase}
\label{subsubsec_disordered} 

\begin{figure} 
\centerline{\includegraphics{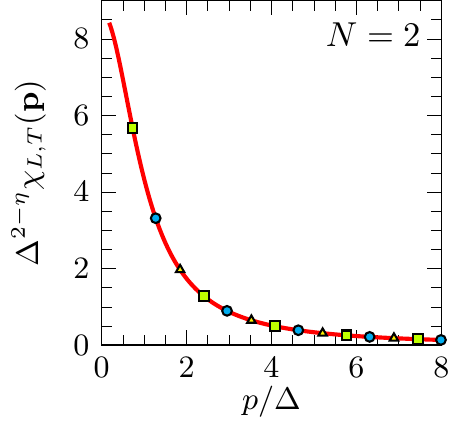}
\includegraphics{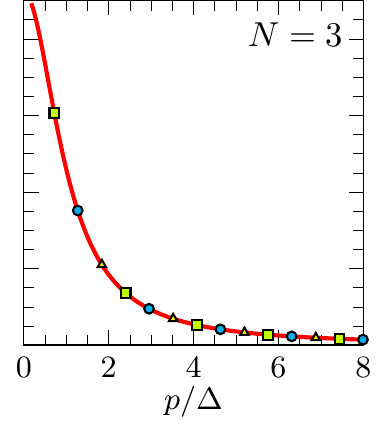}}
\vspace{0.25cm}
\centerline{\includegraphics{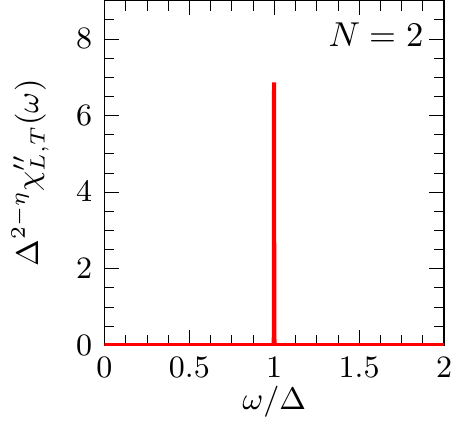}
\includegraphics{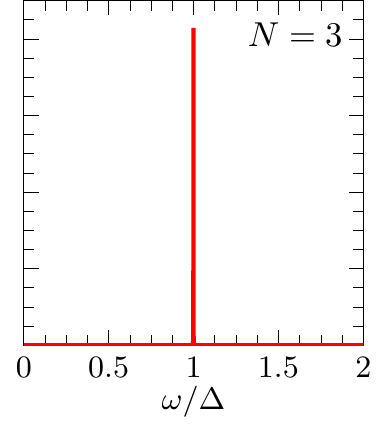}}
\vspace{-0.25cm}
\caption{(Color online) $\chi_{\rm L,T}(\p)$ and $\chi''_{\rm L,T}(\w)$ in the disordered phase for $N=2$  and $N=3$. The solid line and the symbols correspond to different values of $r_0-r_{0c}$.}
\label{fig_chiL_disordered} 
\centerline{\includegraphics{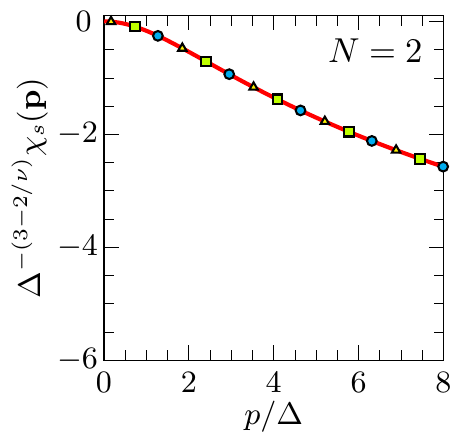}
\includegraphics{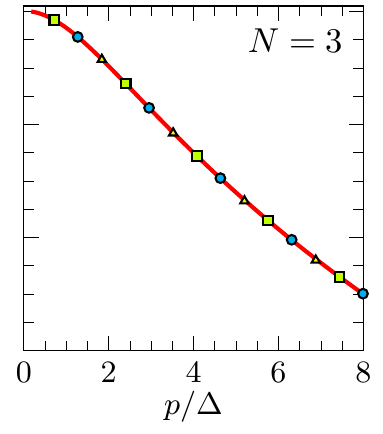}}
\vspace{0.25cm}
\centerline{\includegraphics{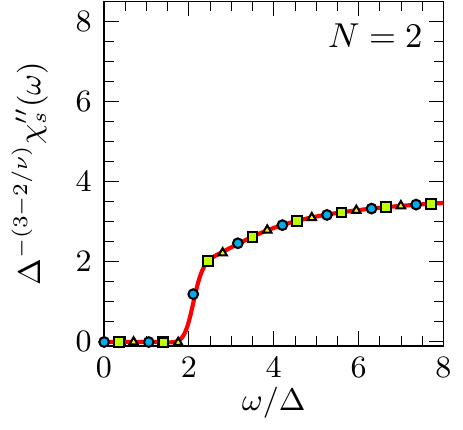}
\includegraphics{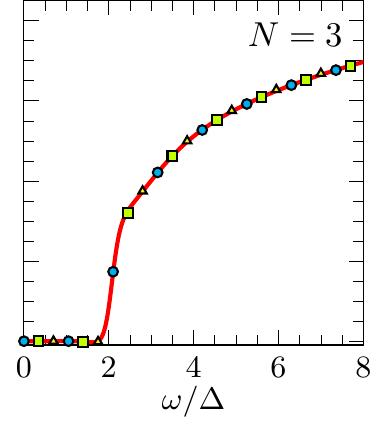}}
\vspace{-0.25cm}
\caption{(Color online) Same as Fig.~\ref{fig_chiL_disordered} but for the scalar susceptibility $\chi_s$.} 
\label{fig_chis_disordered}
\end{figure}

Figures~\ref{fig_chiL_disordered} and \ref{fig_chis_disordered} show $\chi_{\rm L,T}(\p)$ and $\chi_s(\p)$ and their spectral functions in the disordered phase for $N=2$ and $N=3$. The various curves, obtained for different values of $r_0-r_{0c}$, show a data collapse in agreement with the scaling forms~(\ref{chiLT_scaling},\ref{chis_scaling}) expected in the critical regime. The excitation gap $\Delta$, deduced from the peak in the spectral function $\chi_{\rm L,T}''(\w)$, is in very good agreement with the approximate expression~(\ref{Deltaest}). 

The spectral function $\chi_s''(\w)$ of the scalar susceptibility vanishes for $|\w|<2\Delta$. Contrary to previous conclusions based on QMC and NPRG,\cite{Pollet12,Chen13,Rancon14} we find that $\chi_s''(\w)$ rises smoothly above the threshold at $\w=2\Delta$ with no sign of a local maximum for $\w\gtrsim 2\Delta$. The authors of Ref.~\onlinecite{Gazit13a} argued that in spite of the maximum observed above the threshold in their MC simulations, there is inclusive evidence for a resonance at finite frequency in the disordered phase (the peak carries a small spectral weight and its position is not very robust). We also note that no resonance is obtained in the $4-(d+1)$ expansion.\cite{Katan15}

\subsubsection{Ordered phase}  
\label{subsubsec_ordered}

\begin{figure}
\centerline{\includegraphics{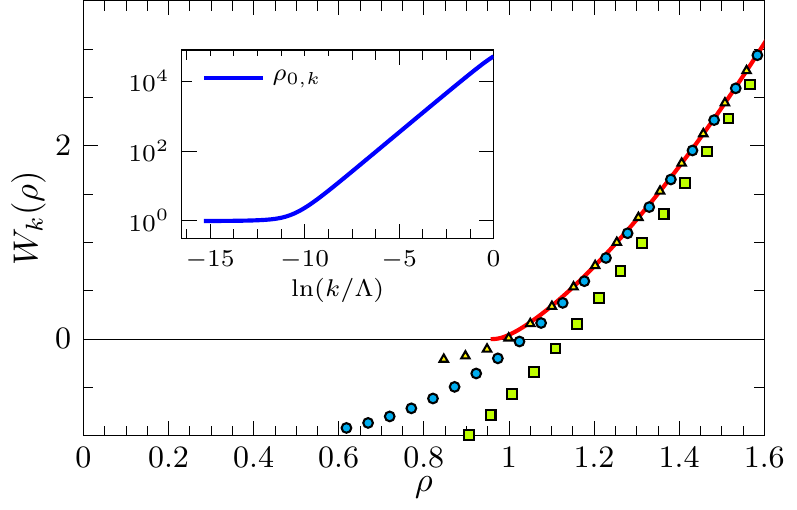}}
\caption{(Color online) Derivative $W_k(\rho)$ of the effective potential, shown in the range $[\rho_{{\rm min},k},\rho_{\rm max}]$, for various values of $k$ (from bottom to top: $\ln(k/\Lambda)=-11.5$, $-12.25$, $-13$ and $-15.25$) and $N=2$. The inset shows the corresponding flow of $\rho_{0,k}$ vs $\ln(k/\Lambda)$.}
\label{fig_Wk} 
\end{figure}

In Fig.~\ref{fig_Wk}, we show the derivative $W_k(\rho)$ of the effective potential for various values of $k$ and $N=2$. As explained above, for small $k$ we must use a $k$-dependent grid $[\rho_{{\rm min},k},\rho_{\rm max}]$ to get rid of the smallest $\rho$ values for which the propagator is not positive. For $k=k_{\rm min}\simeq 0.05\Delta$, $\rho_{\rm min,k}$ is very close to $\rho_{0,k}$ and we cannot continue the flow. In Fig.~\ref{fig_Wk} we also show the behavior of $\rho_{0,k}$ and its convergence towards its $k=0$ value. The extrapolated value at $k=0$ differs from the value at $k_{\rm min}$ by less than 1$\%$. 

\begin{table}
\begin{center}
\caption{Universal ratio $\rho_s/(N\Delta)$ obtained from the NPRG in the BMW approximation (NPRG BMW). Also shown are the previous NPRG results obtained from a derivative expansion of the effective action\cite{Rancon13a} (NPRG DE) and from Monte Carlo simulations\cite{Gazit13a} (MC). The exact result in the limit $N\to\infty$ is $1/4\pi\simeq 0.0796$.}
\begin{tabular}{ccccccccc}
\hline \hline
$N$ & 1000 &  10 & 8 & 6 & 4 & 3 & 2 \\
\hline
NPRG BMW  & 0.0796 &  0.0803 & 0.0829  & 0.0903  & 0.111 & 0.137 & 0.193 \\
NPRG DE\cite{Rancon13a} & 0.0838 & 0.085 & 0.086 & 0.089 & 0.096 & 0.106 & 0.132  \\
MC\cite{Gazit13a} & & &   &  & & 0.114 & 0.220  \\
\hline \hline
\end{tabular}
\label{table_rhos}
\end{center}
\end{table}

Table~\ref{table_rhos} shows the universal ratio $\rho_s/(N\Delta)$ where $\rho_s$ and $\Delta$ are computed for the same distance $|r_0-r_{0c}|$ to the QCP (see the discussion at the end of Sec.~\ref{subsec_gamk}). For small values of $N$ we find significant deviations wrt previous NPRG results.\cite{Rancon13a} The value for $N=2$ is now much closer to the MC estimate of Ref.~\onlinecite{Gazit13a} and the agreement is also satisfactory for $N=3$. For $N=1000$, we recover the large-$N$ result. 

\begin{figure} 
\centerline{\includegraphics{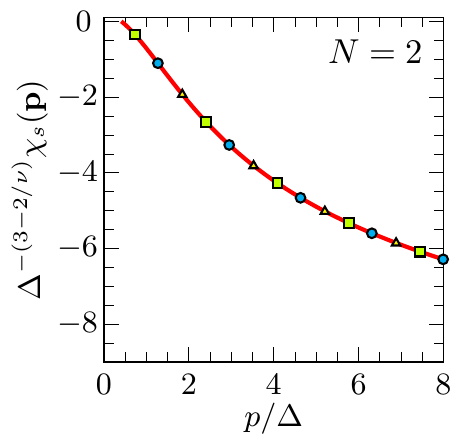}
\includegraphics{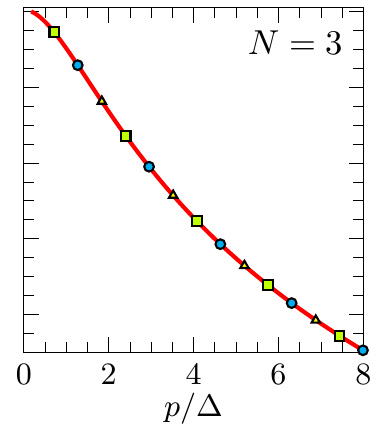}}
\vspace{0.25cm}
\centerline{\includegraphics{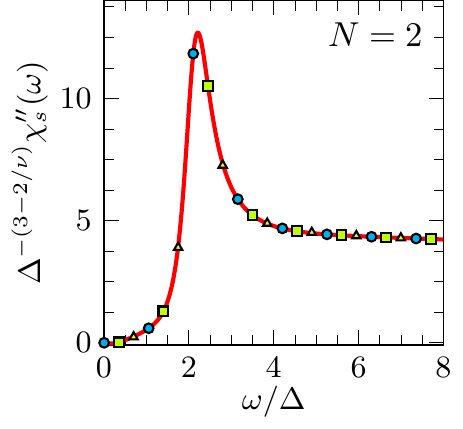}
\includegraphics{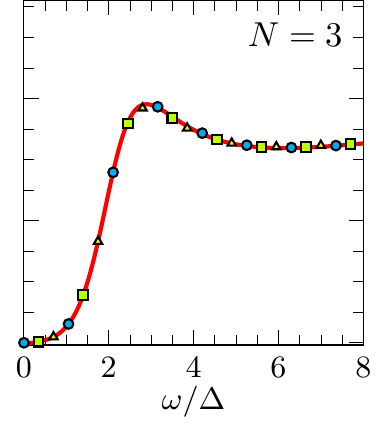}}
\caption{(Color online) $\chi_s(\p)$ and $\chi''_s(\w)$ in the ordered phase for $N=2$ and $N=3$. The solid line and the symbols correspond to different values of $r_0-r_{0c}$.}
\label{fig_chis_ordered} 
\vspace{0.25cm}
\centerline{\includegraphics{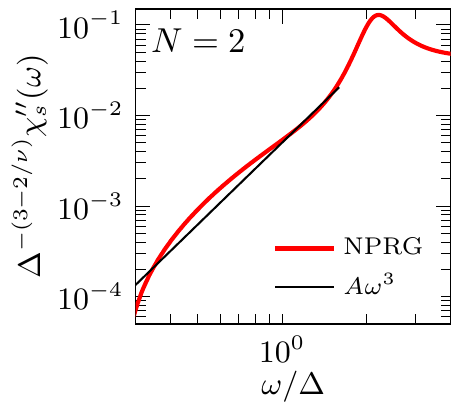}
\includegraphics{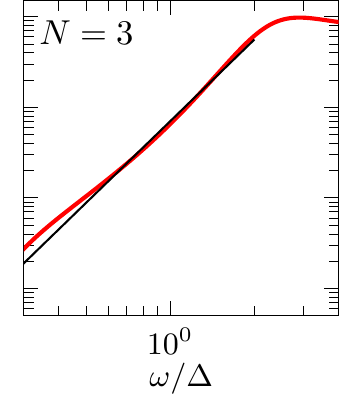}}
\caption{(Color online) Log-log scale plot of $\chi''_{s}(\w)$ in the ordered phase for $N=2$ and $N=3$, showing the asymptotic behavior $\chi''_{s}(\w)\sim \w^3$ at low energies.}
\label{fig_chis_smallwscaling}
\end{figure}

\begin{table}
\begin{center}
\caption{Universal ratio $m_H/\Delta$ obtained from the NPRG in the BMW approximation. Also shown are previous NPRG results\cite{Rancon14} as well as results obtained from (Q)MC\cite{Gazit13,Chen13} and $\eps$ expansion.\cite{Katan15}}
\begin{tabular}{ccc}
\hline \hline
$N$ & 3 & 2 \\
\hline
NPRG BMW & 2.7 & 2.2  \\
NPRG\cite{Rancon14} &  & 2.4 \\
MC\cite{Gazit13} & 2.2(3) & 2.1(3)  \\
QMC\cite{Chen13} &  & 3.3(8)  \\
$\eps$ expansion\cite{Katan15} & 1.64 & 1.67 \\
\hline \hline
\end{tabular}
\label{table_mH}
\end{center}
\end{table}

\begin{figure} 
\centerline{\includegraphics{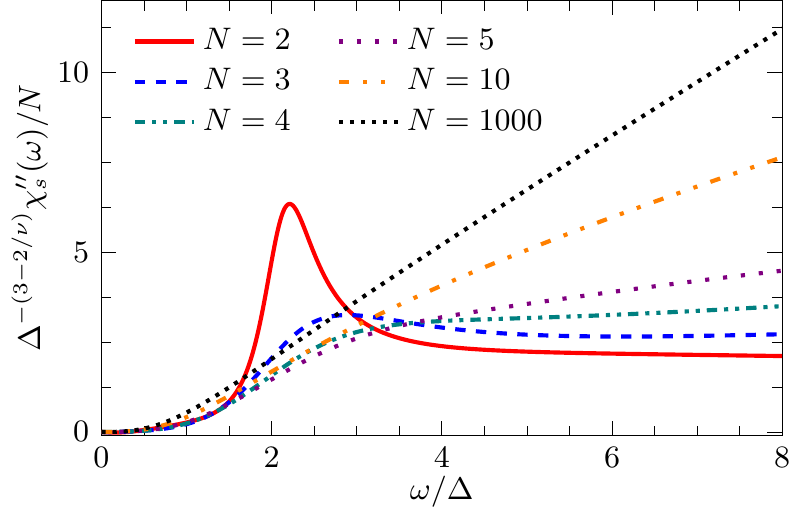}}
\caption{(Color online) Spectral function $\chi_s''(\w)$ for various values of $N$ in the ordered phase.} 
\label{fig_chis_ordered_allN}
\end{figure}

In Fig.~\ref{fig_chis_ordered} we show $\chi_s(\p)$ and $\chi_s''(\w)$ in the ordered phase for $N=2$ and $N=3$. Again, we observe data collapse in agreement with the scaling forms~(\ref{chis_scaling}). For $N=2$, we find a well-defined Higgs resonance whose position $\w=m_H$ and full width at half-maximum vanishes as the QCP is approached. For $m_H\ll \w\ll p_G$, we recover the critical scaling $\chi_s''(\w)\sim \w^{3-2/\nu}$. Up to a multiplicative factor which depends on the nonuniversal factor $\calA_-$ [Eq.~(\ref{chis_scaling})] the shape of the resonance, given by the universal scaling function $\Phi_{s,-}$, is in very good agreement with the MC result of Refs.~\onlinecite{Gazit13,Gazit13a}. The Higgs resonance is still visible, although less pronounced, for $N=3$. This observation disagrees with previous NPRG
results\cite{Rancon14} but agrees with MC simulations of Ref.~\onlinecite{Gazit13a}. The universal ratio $m_H/\Delta$, shown in Table~\ref{table_mH}, is compatible with MC estimates of Refs.~\onlinecite{Gazit13,Gazit13a}. Since in the ordered phase we must stop the flow at a finite value $k_{\rm min}$, we cannot calculate reliably the spectral function $\chi_s''(\w)$ for frequencies $\w\lesssim k_{\rm min}$. Although for $k_{\rm min}\lesssim\w\lesssim m_H$, our results are compatible with $\chi_s''(\w)\sim\w^3$ (see Fig.~\ref{fig_chis_smallwscaling}), the low-energy regime $\w\ll\Delta$ where the spectral function is completely determined by the Goldstone modes is difficult to access. In Fig.~\ref{fig_chis_ordered_allN} we show $\chi_s''(\w)$ for $N=2,3,4,5,10,100$. Only for $N=2$ and $N=3$ does a Higgs resonance exist.

\begin{figure}
\centerline{\includegraphics{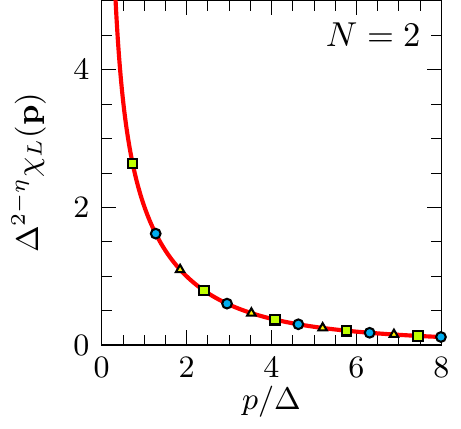}
\includegraphics{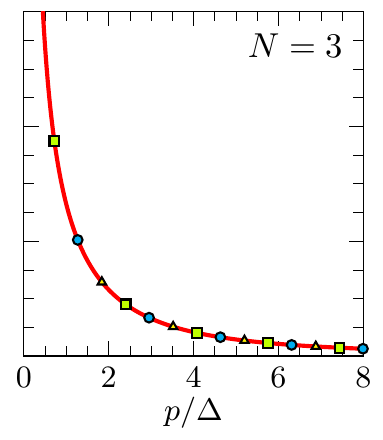}}
\vspace{0.25cm}
\centerline{\includegraphics{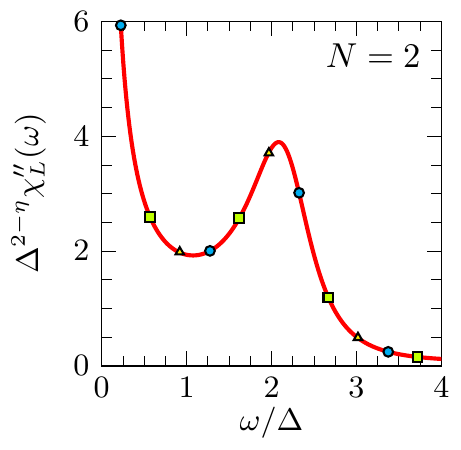}
\includegraphics{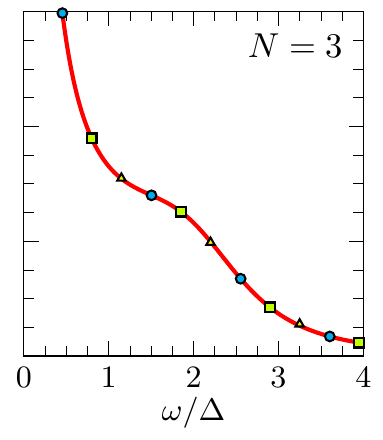}}
\caption{(Color online) $\chi_{\rm L}(\p)$ and $\chi''_{\rm L}(\w)$ in the ordered phase for $N=2$ and $N=3$. The solid line and the symbols correspond to different values of $r_0-r_{0c}$.} 
\label{fig_chiL_ordered}
\centerline{\includegraphics{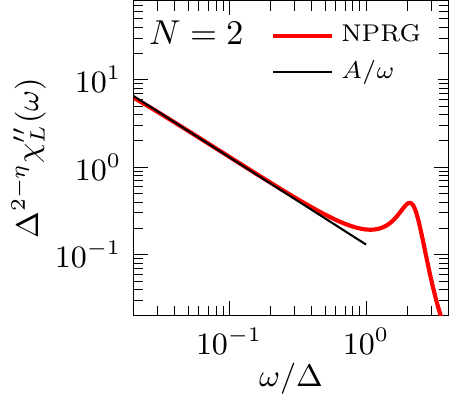}
\includegraphics{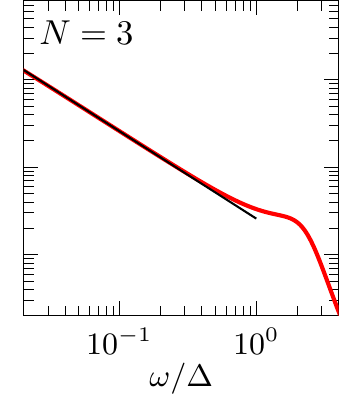}}
\caption{(Color online) Log-log scale plot of $\chi''_{\rm L}(\w)$ in the ordered phase for $N=2$ and $N=3$, showing the asymptotic behavior $\chi''_{\rm L}(\w)\sim 1/\w$ at low energies.\cite{not3}}
\label{fig_chiL_smallwscaling}
\end{figure}

Finally we show the longitudinal susceptibility $\chi_{\rm L}(\p)$ and its spectral function $\chi''_{\rm L}(\w)$ in Fig.~\ref{fig_chiL_ordered} for $N=2$ and $N=3$. For $\p\to 0$, the longitudinal susceptibility $\chi_{\rm L}(\p)$ diverges as $1/p$ as expected for a two-dimensional system (Fig.~\ref{fig_chiL_smallwscaling}).\cite{not3} 
This effect is a consequence of the coupling of the longitudinal mode to the Goldstone modes\cite{Patasinskij73,Sachdev99,Zwerger04,Dupuis11} and prevents the observation of a well-defined Higgs resonance in $\chi_{\rm L}''(\w)$.\cite{Podolsky11} Nevertheless a broad peak, presumably due to the Higgs mode, can be seen for $\w\sim m_H$ when $N=2$ (Fig.~\ref{fig_chiL_ordered}).\cite{not2} For $N=3$, the peak has disappeared but a faint structure can still be seen. 

\section{Conclusion} 
\label{sec_conlu}

We have studied the scalar and longitudinal susceptibilities in the quantum O($N$) model using the NPRG. Comparison with QMC simulations\cite{Pollet12,Chen13,Gazit13,Gazit13a} and $\eps=4-(d+1)$ expansion\cite{Katan15} allows us to identify robust properties of the Higgs mode: 
i) In the ordered phase, there is a well-defined Higgs resonance for $N=2$ and $N=3$. The spectral function $\chi_s''(\w)$ has been determined both from QMC and NPRG but the precise value of the the mass of the Higgs mode is not precisely known (Table~\ref{table_mH}). If we take the difference between NPRG and MC simulations of Refs.~\onlinecite{Gazit13,Gazit13a} 
as an estimate of the error, then the ratio $m_H/\Delta$ is known within 5$\%$ for $N=2$ and $20\%$ for $N=3$. ii) In the disordered phase, there is no Higgs-like peak in $\chi_s''(\w)$ above the absorption threshold. 
There are two other important properties obtained from the NPRG that have not been studied with MC or other methods so far: iii) The Higgs resonance is 
suppressed for $N\geq 4$.  iv) For $N=2$ the Higgs mode manifests itself in the longitudinal spectral function $\chi''_{\rm L}(\w)$ by a very broad peak.

From a more technical point of view, we have shown that the BMW approximation\cite{Blaizot06,Benitez09,Benitez12,Parola84,Parola95} allows one to compute the momentum dependence of correlation functions, including 4-point functions such as the scalar susceptibility. We have also shown that the difficulties arising from the non-positivity of the propagator in the ordered phase can be overcome by using a $k$-dependent grid which does not include small values of the order parameter $\rho$.

\begin{acknowledgments}
We thank N. Wschebor for a critical reading of the manuscript. 
ND thanks A. Ran\c{c}on for a previous collaboration and for suggesting the large-$N$ calculation of Appendix~\ref{app_largeN}. 
Universit\'e Pierre et Marie Curie is part of Sorbonne Universit\'es. 
\end{acknowledgments} 

\appendix

\section{Vertices $\Gamma_k^{(n,m)}$ in the large-$N$ limit}
\label{app_largeN} 

In this section, we determine the effective potential $W_k$ and the vertices $\Gamma_k^{(2,0)}$, $\Gamma_k^{(0,2)}$ and $\Gamma_k^{(1,1)}$ in the large-$N$ limit using the standard approach where the partition function $Z_k[h]\equiv Z_k[h,\J=0]$ is obtained from a saddle-point calculation.\cite{Zinn_book_2,Dupuis11} We first introduce the field $\rho=\varphibf^2$ and a Lagrange multiplier $\lamb$,
\begin{widetext}
\begin{align}
Z_k[h] &= \int \calD[\varphibf,\rho,\lambda] \exp \biggl\lbrace -\int_\x \Bigl[ \half(\nablabf\varphibf)^2+ \left(\frac{r_0}{2}-h\right) \rho + \frac{u_0}{4!N} \rho^2 + i \frac{\lambda}{2} (\varphibf^2-\rho) \Bigr] 
- \half \int_{\x,\x'} \varphibf(\x) R_k(\x-\x') \cdot \varphibf(\x') \biggr\rbrace \nonumber \\ 
&= \int \calD[\varphibf,\lambda] \exp\biggl\lbrace \int_\x \biggl[ \frac{3N}{2u_0} (2h+i\lambda-r_0)^2
-\half \left[ (\nablabf\varphibf)^2 + i \lambda \varphibf^2 \right] \biggr] 
- \half \int_{\x,\x'} \varphibf(\x) R_k(\x-\x') \cdot \varphibf(\x') \biggr\rbrace .
\end{align}
Then we split the field $\varphibf$ into a $\sig$ field and an $(N-1)$-component field $\pibf$. Integrating over the $\pibf$ field, we obtain the action 
\begin{equation}
S_k[\sig,\lamb,h] = \int_\x \biggl[ -\frac{3N}{2u_0} (2h+i\lambda-r_0)^2
+ \half \left[ (\nablabf\sig)^2 + i \lambda \sig^2 \right]  \biggr]
+  \half \int_{\x,\x'} \sig(\x) R_k(\x-\x') \sig(\x') 
+ \frac{N-1}{2} \Tr \ln g_k^{-1}[\lamb] , 
\end{equation}  
\end{widetext} 
where 
\begin{equation} 
g_k^{-1}[\x,\x';\lamb] = [-\nablabf^2 + i\lamb(\x) ] \delta(\x-\x') + R_k(\x-\x')
\end{equation}
is the inverse propagator of the field $\pi_i$ in the fluctuating $\lamb$ field. In the limit $N\to\infty$, the action becomes proportional to $N$ (if one rescales the $\sig$ field, $\sig\to\sqrt{N}\sig$); the saddle point approximation becomes exact for the partition function $Z_k[h]$ and the Legendre transform of the free energy coincides with the action $S_k$.\cite{Lebellac_book} This implies that the scale-dependent effective action~(\ref{gammak}), defined as the Legendre transform including the subtraction of $\Delta S_k[\sig]$, is simply equal to $S_k[\sig,\lamb,h]-\Delta S_k[\sig]$:
\begin{multline}
\Gamma_k[\sig,\lamb,h] = \int_\x \biggl\lbrace -\frac{3N}{2u_0} (2h+i\lambda-r_0)^2 \\
+ \half \left[ (\nablabf\sig)^2 + i \lambda \sig^2 \right] \biggr\rbrace 
+ \frac{N}{2} \Tr \ln g_k^{-1}[\lamb] 
\end{multline} 
(we use $N-1\simeq N$ for large $N$). We can eliminate the Lagrange multiplier field using 
\begin{equation}
\frac{\delta \Gamma_k[\sig,\lamb,h]}{\delta\lamb(\x)} \biggl|_{\lamb=\lamb_k[\sig,h]} = 0 ,
\label{app1}
\end{equation}
to obtain a scale-dependent effective action $\Gamma_k[\sig,h]\equiv\Gamma_k[\sig,\lamb_k[\sig,h],h]$ which is a functional of $\sig$ and $h$. $\Gamma_k[\sig,h]$ is the starting point to compute the vertices $\Gamma_k^{(n,m)}$ in the large-$N$ limit.

Let us first consider the effective potential for $h=0$, 
\begin{align}
U_k(\rho) ={}& - \frac{3N}{2u_0} (i\lamb_k-r_0)^2 + i \lamb_k \rho \nonumber \\ & 
+ \frac{N}{2} \int_\q \ln[\q^2+i\lamb_k+R_k(\q)] ,
\end{align}
where we use the notation $\rho=\sig^2/2$. The value of $\lamb_k\equiv\lamb_k(\rho)$ is obtained from $\partial U_k/\partial\lamb_k=0$, which follows from~(\ref{app1}), i.e. 
\begin{equation}
- \frac{3N}{u_0} (i\lamb_k-r_0) + \rho + \frac{N}{2} \int_\q \frac{1}{\q^2+i\lamb_k+R_k(\q)} = 0. 
\end{equation}
We deduce that 
\begin{align} 
W_k(\rho) &= i\lamb_k \nonumber \\ 
&= r_0 + \frac{u_0\rho}{3N} + \frac{u_0}{6} \int_\q \frac{1}{\q^2+W_k(\rho)+R_k(\q)} ,
\label{app4} 
\end{align} 
which is the known result in the limit $N\to\infty$.\cite{Zinn_book_2}

The vertex $\Gamma_k^{(2,0)}$ can be obtained from $\Gamma_k[\sig,\lamb]$ setting $h=0$. In Fourier space, the 2-point vertex $\Gamma_k^{(2)}$ can be written as a $2\times 2$ matrix with components $\Gamma^{(2)}_{k,\sig\sig}$, $\Gamma^{(2)}_{k,\sig\lamb}$, $\Gamma^{(2)}_{k,\lamb\sig}$ and $\Gamma^{(2)}_{k,\lamb\lamb}$.\cite{Dupuis11} Inverting this matrix, one obtains the longitudinal propagator as $(\Gamma_k^{(2)-1})_{\sig\sig}(\p)\equiv[\GamA(\p)+2\rho\GamB(\p)]^{-1}$ from which we deduce (see Eq.~(55) in Ref.~\onlinecite{Dupuis11}) 
\begin{equation}
\begin{gathered}
\Gamma_{A,k}(\p,\rho) = \p^2, \\
\Gamma_{B,k}(\p,\rho) = \left[ \frac{3N}{u_0} + \frac{N}{2} \Pi_k(\p,\rho) \right]^{-1} . 
\end{gathered} 
\end{equation} 
The momentum dependence of the transverse 2-point vertex $\Gamma_{k,\rm T}(\p,\rho)=\p^2+W_k(\rho)$ remains the bare one. 

Let us now consider the vertex $\Gamma_k^{(0,2)}(\x,\x',\rho)$, 
\begin{equation}
\Gamma_k^{(0,2)}(\x,\x',\rho) = \frac{\bar\delta^2 \Gamma_k[\sig,\lamb_k[\sig,h],h]}{\bar\delta h(\x) \bar\delta h(\x')} \biggl|_{\sig(\z)=\sig,h=0} ,
\end{equation}
where $\sig=\sqrt{2\rho}$ and $\bar\delta/\bar\delta h(\x)$ denotes a total derivative (Sec.~\ref{subsec_chis}). Using~(\ref{app1}), we obtain 
\begin{multline}
\Gamma_k^{(0,2)}(\x,\x',\rho) = \frac{\delta^2 \Gamma_k[\sig,\lamb_k[\sig,h],h]}{\delta h(\x)\delta h(\x')} \biggl|_{\sig(\z)=\sig,h=0} \\
+ \int_\y \frac{\delta^2 \Gamma_k[\sig,\lamb,h]}{\delta h(\x)\delta\lamb(\y)} 
\frac{\delta \lamb_k[\y;\sig,h]}{\delta h(\x')} \biggl|_{\sig(\z)=\sig,\lamb=\lamb_k[\sig,0],h=0} 
\end{multline} 
and in turn 
\begin{align}
\Gamma_k^{(0,2)}(\x,\x',\rho) ={}& - \frac{12N}{u_0} \delta(\x-\x') \nonumber \\ & 
- i \frac{6N}{u_0} \frac{\delta\lamb_k[\x;\sig,h]}{\delta h(\x')} 
\biggl|_{\sig(\z)=\sig,h=0} .
\label{app2} 
\end{align} 
To determine the last term of~(\ref{app2}), we take the functional derivative $\bar\delta/\bar\delta h(\x')$ of Eq.~(\ref{app1}), which gives
\begin{align} 
0 ={}& -i \frac{6N}{u_0} \delta(\x-\x') \nonumber \\ & 
+ \int_\y \Gamma^{(2)}_{k,\lamb\lamb}(\x,\y,\rho) \frac{\delta\lamb_k[\y;\sig,h]}{\delta h(\x')} \biggl|_{\sig(\z)=\sig,h=0} ,
\label{app3} 
\end{align} 
where 
\begin{equation}
\Gamma^{(2)}_{k,\lamb\lamb}(\x,\x',\rho) = \frac{\delta^2 \Gamma_k[\sig,\lamb,h]}{\delta\lamb(\x) \delta\lamb(\x')} \biggl|_{\sig(\z)=\sig,\lamb=\lamb_k[\sig,0],h=0} .
\end{equation}
From~(\ref{app2},\ref{app3}), we finally deduce 
\begin{equation}
\Gamma_k^{(0,2)}(\p,\rho) = - \frac{12N}{u_0} + \left(\frac{6N}{u_0}\right)^2 \Gamma^{(2)}_{k,\lamb\lamb}(\p,\rho)^{-1} , 
\label{app5}
\end{equation}
where 
\begin{equation}
\begin{gathered} 
\Gamma^{(2)}_{k,\lamb\lamb}(\p,\rho) = \frac{3N}{u_0} + \frac{N}{2} \Pi_k(\p,\rho) , \\
\Pi_k(\p,\rho) = \int_\q g_k(\q,\rho) g_k(\p+\q,\rho)  
\end{gathered}
\end{equation}
and $g_k(\q,\rho) = [\q^2 + W_k(\rho)+R_k(\q)]^{-1}$ (we use $i\lamb_k=W_k(\rho)$).  

Following a similar approach for
\begin{equation}
\Gamma_{k,1}^{(1,1)}(\x,\x',\rho) = \frac{\bar\delta^2 \Gamma_k[\sig,\lamb_k[\sig,h],h]}{\bar\delta\sig(\x) \bar\delta h(\x')} \biggl|_{\sig(\z)=\sig,h=0} ,
\end{equation}
one finds 
\begin{equation}
\Gamma_{k,1}^{(1,1)}(\p,\rho) = - \frac{6N}{u_0} \sqrt{2\rho} \Gamma^{(2)}_{k,\lamb\lamb}(\p,\rho)^{-1} .
\label{app6}
\end{equation} 

%



\begin{thebibliography}{43}
\bibitem{not5} 
The value $\alpha=2.25$ is near the optimal value giving the best estimate of the critical exponents.\cite{Benitez12} %
\bibitem{not4} In practice, we obtain better results for the scalar susceptibility if we use $\gam_k$ rather than $\tilde \gam_k$. This avoids a poor numerical precision in the computation of $\gam_k$  due to the multiplication of a small quantity, $\tilde\gam_k\sim k^2$, by a large one, $k^{-2}$ [See Eq.~(\ref{vdim})].
\bibitem{not7}
The positivity condition $W_k(\rho)+R_k(\p=0)$ is violated when $W_k(\rho)<-\alpha$, i.e.
$W_k(\rho)<-2.25$; we adopt here a stricter condition for numerical reasons.%
\bibitem{not6} The value of $k_{\rm min}$ is not sensitive to the value of the cutoff function $R_k(\q=0)=\alpha Z_kk^2$ (which can be changed by varying $\alpha$). %
\bibitem [{not({\natexlab{a}})}]{not3}%
  \BibitemOpen
  \href@noop \bibinfo  {For $N=2$, the numerical
  fit of $\chi''_{\rm L}(\w)$ gives an exponent $\simeq 0.8$ instead of 1. A
  possible explanation is that the actual infrared behavior $\chi''_{\rm
  L}(\w)\sim 1/\w$ is visible only at lower frequencies where our approach
  fails (see the discussion at the beginning of
  Sec.~\ref{sec_spectral}).}\BibitemShut {Stop}%
\bibitem [{not({\natexlab{b}})}]{not2}%
  \BibitemOpen
  \href@noop \bibinfo {The maximum of the broad
  peak in $\chi''_{\rm L}(\w)$ is actually observed at $\w<m_H$ as
  expected.}\BibitemShut {Stop}%
\end{thebibliography}

\end{document}